\documentclass[prl,reprint,superscriptaddress,longbibliography,floatfix,twocolumns,nofootinbib]{revtex4-1}
\usepackage[utf8]{inputenc}
\usepackage{amsmath,graphicx, dsfont, bm, amsthm, amssymb,color,physics,xcolor}
\usepackage[normalem]{ulem}

\newcommand{\ii}{\mathrm{i}}

\newcommand{\cE}{\mathcal{E}}

\newcommand{\cH}{\mathcal{H}}

\renewcommand{\t}[1]{\mathrm{#1}}
\newcommand{\be}{\begin{equation}}
\newcommand{\ee}{\end{equation}}

\newcommand{\cw}{\circlearrowright}
\newcommand{\acw}{\circlearrowleft}

\begin{document}

\title{Partial self-testing and randomness certification in the triangle network }
\author{Pavel Sekatski}
\email{pavel.sekatski@unige.ch}
\affiliation{Department of Applied Physics, University of Geneva, 1211 Geneva, Switzerland}
\author{Sadra Boreiri}
\affiliation{Department of Applied Physics, University of Geneva, 1211 Geneva, Switzerland}
\author{Nicolas Brunner}
\affiliation{Department of Applied Physics, University of Geneva, 1211 Geneva, Switzerland}
\begin{abstract}
    Quantum nonlocality can be demonstrated without inputs (i.e. each party using a fixed measurement setting) in a network with independent sources. Here we consider this effect on ring networks, and show that the underlying quantum strategy can be partially characterized, or self-tested, from observed correlations. Applying these results to the triangle network allows us to show that the nonlocal distribution of Renou et al. [Phys. Rev. Lett. 123, 140401 (2019)] requires that (i) all sources produce a minimal amount of entanglement, (ii) all local measurements are entangled, and (iii) each local outcome features a minimal entropy. Hence we show that the triangle network allows for genuine network quantum nonlocality and certifiable randomness. 
\end{abstract}
\maketitle

\paragraph{Introduction --}  
Discovered by Bell in the 1960s \cite{Bell}, the phenomenon of quantum nonlocality has been traditionally investigated in a setting where two (or more) separated observers perform local measurements on a shared entangled state \cite{review}. One can then prove, e.g. via Bell inequality violation, that the observed correlations are Bell nonlocal, in the sense that they are incompatible with any physical theory satisfying a natural notion of locality, such as in classical physics. Beyond fundamental aspects, quantum nonlocality is also a strong resource for black-box quantum information processing.

Networks offer an intriguing new platform for exploring quantum nonlocality; see \cite{Tavakoli_Review} for a review. The key novelty is that the network structure features several sources, each distributing entanglement to various subsets of the parties. At each party, quantum joint measurements can be performed, which enable the distribution of strong correlations across the entire network. The main idea behind network nonlocality is to investigate the resulting correlations under the assumption that all sources in the network are independent \cite{Branciard_2010,Branciard_2012}. This assumption leads to a formal definition of classical (or network-local) correlations, which can be viewed as a natural generalization of the notion of Bell locality. Characterizing classical and quantum correlations in such networks is a highly challenging task, see e.g. \cite{Chaves2015,wolfe2019inflation,Weilenmann_2018,wolfe2021inflation,Aberg2020,Ligthart2021}.

A central question in this research area is to uncover novel forms of quantum nonlocal correlations inherent to the network structure. In turn, one would like to characterize such new forms of nonlocality and explore their potential for applications in quantum information processing. Our work brings progress towards this second direction.

In 2012, Fritz~\cite{Fritz_2012} and Branciard et al.~\cite{Branciard_2012} discovered that quantum nonlocality can be demonstrated in networks without the need for measurement inputs, i.e. each party performing a single fixed measurement. The example of Fritz considers a simple triangle network, where each pair of parties is connected via a bipartite source, see Fig. 1. While the construction of Fritz can be viewed as a clever embedding of a standard Bell test in the triangle network (see also \cite{Fraser}), Renou et al.~\cite{Renou_2019} presented a strikingly different instance of quantum nonlocality (referred to as RGB4), which they argued is genuine to the triangle network; see also \cite{Gisin_2019,Renou2022,Renou_rigid,Abiuso2022,Pozas2022}. In order to formalize this intuition, the concepts of genuine network nonlocality~\cite{Supic2022} (GNNL) and full network nonlocality~\cite{Pozas_full} (FNNL) where proposed. The first demonstrates the presence of non-classical joint measurements, while the second witnesses the distribution of entanglement by all sources. However, the initial question of whether the RGB4 distribution (or any other quantum nonlocal distribution without inputs) has GNNL features remained open so far.

\begin{figure}
    \centering
    \includegraphics[width=0.6\columnwidth]{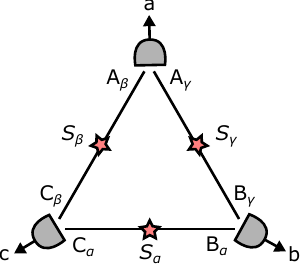}
    \caption{The triangle network features three parties connected pairwise via three independent sources. The figure illustrates the labels we use for sources and subsystems.}
    \label{fig:triangle}
\end{figure}

In this work, we precisely address these questions. We develop methods for the characterization of quantum distributions in networks without inputs. This allows us to partially characterize the RGB4 distribution, and prove the following properties: (i) GNNL, all parties must perform a non-classical measurement, (ii) each source should distribute entanglement, and we obtain a lower bound on the entanglement of formation $\cE_F > 2.5 \%$, and (iii) certified randomness, via a lower bound on the min-entropy $H_{min} > 3.8 \%$. Our main technical results are  self-testing (or quantum rigidity) proofs that apply to quantum (Parity) Token Counting strategies on ring networks. The exposition in the main text will be focused on the triangle, the generalizations are presented at the end.

\paragraph{Triangle network --}  The triangle network depicted in Fig.~\ref{fig:triangle}, involves three parties $A$, $B$ and $C$. Each pair of parties is connected by a bipartite source, labeled with $\alpha, \beta$, and $\gamma$. Each party receives two systems (from the neighboring sources) and produces an output $a,b$, and $c$. There are in total six involved systems labeled $X_\xi$, with $X\in\{A,B,C\}$ referring to the party receiving the system and $\xi\in\{\alpha,\beta,\gamma\}$ to the source preparing it. 

The set of output probability distributions $P(a,b,c)$ possible on the triangle depends on the physical theory used to model the experiment. Classically, a source distributes (correlated) random variables, from which the measurements produce an output. In quantum theory, a Hilbert space  is associated to each system, which for simplicity we assume to have an arbitrarily large but finite dimension. 
Without loss of generality one can assume that the sources distribute pure states  $\ket{\psi_\alpha}_{B_\alpha C_\alpha}$, $\ket{\psi_\beta}_{C_\beta A_\beta}$, $\ket{\psi_\gamma}_{A_\gamma B_\gamma}$, with the global state denoted
\be\label{eq: triangle state}
\ket{\Psi} = \ket{\psi_\alpha}_{B_\alpha C_\alpha} \ket{\psi_\beta}_{C_\beta A_\beta}\ket{\psi_\gamma}_{A_\gamma B_\gamma}.
\ee
The Hilbert space associated to each system is taken as the support of the state, e.g. $\cH_{B_\alpha} = \t{supp}(\tr_{C_\alpha} \ketbra{\psi_\alpha})$. Thus, when discussing a system in the following, we refer to the Hilbert space where $\ketbra{\Psi}$ is supported, which is natural in the device-independent framework. The measurements performed by the  parties are modeled by positive operator valued measures (POVMs) $\{E_A^a\}_a, \{E_B^b\}_b$, $ \{E_C^c\}_c$, by Stinespring's dilation theorem, we can dilate the measurements to projective ones $\{\bar \Pi_A^a\}_a, \{\bar \Pi_B^b\}_b$, $ \{\bar \Pi_C^c\}_c$. This is done by introducing an auxiliary system $M_X$ prepared in the state $\ket{0}_{M_X}$ for each party,  so that the projectors $(\Pi_X^x)^2=\Pi_X^x$ act on systems $A_\beta A_\gamma M_A$, $B_\gamma B_\alpha M_B$ and $C_\alpha C_\beta M_C$ respectively and satisfy $E_X^x=\tr_{M_X} \bar \Pi_X^x \ketbra{0}_{M_X} $, see appendix A for details.  The output probability distribution is given by
\be
P(a,b,c)= \bra{\Psi,\bm 0}\bar \Pi_A^a \bar \Pi_B^b \bar \Pi_C^c\ket{\Psi,\bm 0},
\ee
with $\ket{\bm 0} = \ket{0,0,0}_{M_AM_BM_C}$.

\paragraph{Self-testing Token Counting distributions --}

Let us now focus on a particular family of \textit{classical models} on the triangle, called Token Counting (TC) strategies. In such a strategy a source $\xi$ randomly distributes a fixed number of tokens $N_\xi$ left or right -- with probability $p_\xi(i)$ there are $i$ tokens sent to the left and $(N_\xi-i)$ tokens to the right. Each party then outputs the total number of tokens it received. The resulting correlations $P(a,b,c)$ are called TC distributions. By construction, all such distributions fulfill the constraint
\be\label{eq: TC rigid}
a+b+c=N,
\ee
where $N\equiv N_\alpha+N_\beta+N_\gamma$. The TC distributions are known to be \textit{rigid}~\cite{Renou_rigid}, meaning that among all possible classical strategies on the triangle the TC strategy we just described is essentially the unique model leading to $P(a,b,c)$. We will now show that this result can be generalized to \textit{quantum} strategies.

We first remark that for a quantum model the constraint~\eqref{eq: TC rigid} can be put in the form
\be\label{eq: TC rigid quantum}
 \bar \Pi_A^a \bar \Pi_B^b \bar \Pi_C^c \ket{\Psi,\bm 0} = 0 \quad  \text{if}\quad   a+ b + c \neq N.
\ee
Next, for each party $X$ let us define an unitary operator
\be\label{eq: U_x def}
\bar U_X \equiv \sum_{x} e^{\ii \varphi_x} \bar \Pi_{X}^x, \quad \text{with}\quad  \varphi_x=\frac{2\pi (x+1/3)}{N+1}
\ee
where the integer $x$ runs through possible outputs of the party. These definitions allow us to put the  constraint~\eqref{eq: TC rigid quantum}  in a particularly simple form
\be\label{eq: rigid cond}
\bar U_A \bar U_B \bar U_C \ket{\Psi,\bm 0} = \ket{\Psi,\bm 0}.
\ee
To see this note that $\bar U_A \bar U_B \bar U_C$ is a global unitary with eigenvalues $e^{\ii (a+b+c+1)\frac{2 \pi }{N+1}}$, furthermore Eq.~\eqref{eq: TC rigid quantum} guarantees that the state is only supported on the subspace associated to the eigenvalue $e^{\ii (a+b+c+1)\frac{2 \pi }{N+1}}=1$.  This condition implies that the dilation of measurements is trivial as summarized by the following observation.

\textbf{Result 0.} \textit{For unitaries $\bar U_X$ defined in Eq.~\eqref{eq: U_x def}, the identity~\eqref{eq: rigid cond} implies that the original measurements are projective}
\be
E_{X}^x = \bra{0}_{M_X} \bar \Pi_X^x \ket{0}_{M_X} = \Pi_{X}^x.
\ee
\textit{Proof sketch--} The detailed proof is given in Appendix A. The condition \eqref{eq: rigid cond} ensures that the unitaries do not change the state of the auxiliary systems and imply that the operators $U_X=\bra{0}_{M_X} \bar U_X \ket{0}_{M_X}$ are also unitary. But  $U_X = \sum_x e^{\ii \varphi_x} E_X^{x}$ can only be unitary if $\{E_X^x=\Pi_X^x\}_x$ is a projector valued measure$\square$.

Hence, we can rewrite Eq.~\eqref{eq: rigid cond} in a simpler form $ U_A U_B  U_C \ket{\Psi} =\ket{\Psi}$, where $U_X = \sum_x e^{\ii \varphi_x} \Pi_X^{x}$. This condition implies that the unitaries are product.


\textbf{Result 1.} \textit{Consider a quantum state $\ket{\Psi}=\ket{\psi_\alpha}_{B_\alpha C_\alpha} \ket{\psi_\beta}_{C_\beta A_\beta}\ket{\psi_\gamma}_{A_\gamma B_\gamma}$ on the triangle network, and local unitaries $U_{A},  U_{B}, U_C$ acting on the systems $A_\beta A_\gamma$, $B_\gamma B_\alpha$ and $C_\alpha C_\beta$.  The condition $ U_A  U_B  U_C \ket{\Psi} = \ket{\Psi}$ with $\ket{\bm 0}= \ket{0}_{M_A}\ket{0}_{M_B} \ket{0}_{M_C}$, implies that all the unitaries are product}
\be\nonumber\begin{split}
 U_{A} &= V_{A_\beta}\otimes W_{A_\gamma}\\
U_{B}&= V_{B_\gamma}\otimes W_{B_\alpha} \\
U_{C}&= V_{C_\alpha}\otimes W_{A_\beta}.    
\end{split}
\ee
\textit{with unitary $V_{X_\xi}$ and $W_{X_\xi}$ acting on the respective systems.}

\textit{Proof sketch--}
The proof can be found in Appendix~B for any ring network. We rely on the Schmidt decomposition of the states $\ket{\psi_\xi}$ for the possibility to "move" an operator to act on the other half of an entangled state (upon transposition and rescaling). Together with the Choi-Jamio\l{}kowski isomorphism,  this allows us to express the  constraint~\eqref{eq: rigid cond} as an equality between products of bipartite operators acting on 3 systems. Finally, we prove a technical lemma showing that these operators are products. $\square$

 Next, let us explore the implications of result 1 on the measurements, focusing on Alice.    
The local unitaries can be diagonalized $V_{A_\beta}= \sum_j e^{\ii v_j} \Pi_{A_\beta}^{j}$, $W_{A_\gamma}= \sum_\ell e^{\ii w_\ell} \Pi_{A_\gamma}^\ell$. Result 1 guarantees that $\sum_a e^{\ii \varphi_a}\Pi_A^a = \sum_{i,\ell} e^{\ii(v_j+w_\ell)} \Pi_{A_\beta}^j \Pi_{A_\gamma}^\ell$. Since the eigenvalues have to match,i. e. $e^{\ii(v_j+w_\ell)}=e^{\ii \varphi_a}$ for some a, it is easy to see that $e^{\ii v_j}$ and $e^{\ii w_\ell}$ may take at most $N+1$ different values each.
Hence, the Hilbert space associated to the system $A_\beta$ (or $A_\gamma$) can be split as a direct sum $\cH_{A_\beta}= \bigoplus_{j=0}^N\cH_{A_\beta}^{(j)}$
of subspaces $\cH_{A_\beta}^{(j)}$ on which the different $\Pi_{A_\beta}^j$ project. As it is common in self-testing, one can add enough virtual levels to rewrite the direct sum  as a tensor product $\cH_{A_\beta} =  \mathds{C}^{N+1}_{\textbf{A}_\beta}\otimes \cH_{J_{A\beta}}$
with
$\Pi^j_{A_\beta}=\ketbra{j}_{\textbf{A}_\beta}\otimes \mathds{1}_{J_{A\beta}}$.
This decomposes the system $A_\beta$ into a qudit $\textbf{A}_\beta$  and a ''junk system'' $J_{A\beta}$ on which the measurements act trivially. The same decomposition can be derived for each system and imposes the following form on any quantum model fulfilling Eq.~\eqref{eq: TC rigid quantum}
\begin{align}
\label{eq: decomp PVM}
 E_A^a &= \left(\sum_{(j,\ell)\in \mathds{S}(a)} \ketbra{j}_{\textbf{A}_\beta} \otimes\ketbra{\ell}_{\textbf{A}_\gamma}\right)\otimes \mathds{1}_{J_{A\beta}J_{A\gamma}},\\
 \label{eq: decomp state}
\ket{\psi_\alpha}&= \sum_{i,j=0}^N \Psi_{ij}^{(\alpha)}\ket{ij}_{\textbf{B}_\alpha \textbf{C}_\alpha} \ket{j_\alpha^{(ij)}}_{J_{B\alpha}J_{C\alpha}}.
\end{align}
Here, the unknown states of the junk may in particular contain a copy of the qudit states $\ket{j_\alpha^{(ij)}}=\ket{ij}$ and remain inside the source. In this case, the quantum model becomes classical, once the junk systems are traced out. Finally, with the help of the rigidity result~\cite{Renou_rigid} for classical TC strategies, we arrive at the following result. 

\textbf{Result 2.} \textit{Consider a quantum strategy on the triangle with the global state $\ket{\Psi}=\ket{\psi_\alpha}_{B_\alpha C_\alpha} \ket{\psi_\beta}_{C_\beta A_\beta}\ket{\psi_\gamma}_{A_\gamma B_\gamma}$ and the measurements $\{E_A^a\}_a$,$\{E_B^b\}_b$,$\{E_C^c\}_c$  acting on systems $A_\beta A_\gamma$, $B_\gamma B_\alpha$ and $C_\alpha C_\beta$. If the strategy leads to a TC distribution $P(a,b,c)$, arising from a TC strategy with the $N_\alpha,N_\beta,N_\gamma$ tokens distributed by the sources accordingly to probabilities $p_\alpha(i),p_\beta(j),p_\gamma(k)$, then each quantum system }$X_\xi=\textbf{X}_\xi J_{X\xi}$ \textit{can be decomposed in subsystems }$\textbf{X}_\xi$ \textit{and $J_{X,\xi}$ such that the quantum strategy takes the form}
\be \label{eq: TC quantum thrm triangle}\begin{split}
E_X^x &= \left(\sum_{j+\ell=x} \ketbra{j}_{\textbf{X}_\xi} \otimes \ketbra{\ell}_{\textbf{X}_{\xi'}}\right)\otimes \mathds{1}_{J_{X\xi}J_{X\xi'}} \\
\ket{\psi_\xi}_{X_\xi Y_\xi}&=\sum_{i=0}^{N_\xi} \sqrt{p_\xi(i)} \ket{i,N_\xi-i}_{\textbf{X}_\xi\textbf{Y}_\xi}\ket{j^{(i)}_\xi}_{J_{X\xi}J_{Y\xi}},
\end{split}
\ee
\textit{where $\xi$ and $\xi'$ denote the sources connected to the party $X$, and  $X$ and $Y$ denote the parties connected to the source $\xi$.}

\textit{Proof sketch --}
The full proof can be found in Appendix~C for any ring network. The idea is to observe that any quantum strategy given by Eqs.~(\ref{eq: decomp PVM},\ref{eq: decomp state}) defines a unique classical strategy,  where each source $\xi$ samples integer local variables $(i,j)$ accordingly to the probability distribution $|\Psi_{ij}^{(\xi)}|^2$, and sends them to the neighbouring parties $X_{\xi}$ and $Y_\xi$. Upon receiving two such variables from the neighbouring sources, each party outputs $x(j,\ell)$ for which $(j,\ell)\in \mathds{S}(x)$ in Eq.~\eqref{eq: decomp PVM}. Classical rigidity of TC distributions implies a unique possible $|\Psi_{ij}^{(\xi)}|^2$ and enforces Eq.~\eqref{eq: TC quantum thrm triangle}. $\square$

To illustrate the power of the self-testing (or quantum rigidity) provided by Result 2, we now consider a concrete example of a nonlocal quantum distribution on the triangle.

\smallskip 

\paragraph{RGB4 distribution --} In \cite{Renou_2019} a family of tripartite distribution $P_Q(a,b,c)$ with four-value outcomes $a,b,c \in \{0,2,1_0,1_1\}$ has been proposed. It results from quantum models on the triangle network, where each source distributes the same maximally entangled two qubit state $ \ket{\psi^+} = \frac{1}{\sqrt{2}}\left(\ket{01}+\ket{10}\right)$ and each party performs the same two-qubit projective measurement $\{\Pi^0=\ketbra{00},\Pi^2=\ketbra{11}, \Pi^{1_0} = \ketbra{\bar {1}_0},\Pi^{1_1} = \ketbra{\bar {1}_1}\}$, with
$\ket{ \bar 1_i} = u_i \ket{01}+ v_i \ket{10}$. The values $u_i,v_j$ are given by $u_0=-v_1=\cos(\theta)$ and $v_0= u_1= \sin(\theta)$ for a parameter $\theta \in [0,\pi/4]$.
The resulting distributions, which we call RGB4, are given by
\be\label{eq: RGB$ def}\begin{split}
&P_Q( 1_i, 1_j , 1_k) = \frac{1}{8} (u_i u_j u_k + v_i v_j v_k)^2\\
&P_Q(1_i, 0,2)=\frac{1}{8} u_i^2, \quad
P_Q(1_i, 2,0)= \frac{1}{8} v_i^2 \quad \circlearrowright
\end{split}
\ee
where $\circlearrowright$ signifies that the equation is valid up to cyclic permutations of the parties. All the other 
probabilities $P_Q(a,b,c)$ are strictly zero. 

Interestingly, if the outputs are coarse-grained by merging $1_0$ and $1_1$ into a single outcome $1$, the resulting distribution $\bar{P}_Q(a,b,c)$ with $a,b,c =\{0,1,2\}$ becomes TC (with a single token sent left or right at random). 
Thus by Result 2 we know that the states and the measurement are of the form 
\begin{align}\label{eq: rigid states}
 \ket{\psi_\xi}&= \frac{1}{\sqrt{2}}\big( \ket{01}_{\textbf{X}_\xi \textbf{Y}_\xi}\!\ket{j^{c}_\xi}_{J_\xi}\!\!+
\ket{10}_{\textbf{X}_\xi \textbf{Y}_\xi}\ket{j^{a}_\xi}_{J_\xi}
 \big)
 \\
\Pi_X^0 &= \ketbra{00}_{\textbf{X}_\xi \textbf{X}_{\xi'}}\otimes \mathds{1}_{J_X} \nonumber \\
\Pi_X^2&=\ketbra{11}_{\textbf{X}_\xi \textbf{X}_{\xi'}}\otimes \mathds{1}_{J_X} \label{eq: rigid POVMS}\\
 \Pi_X^1&=\Pi_X^{1_0}+\Pi_X^{1_1}=(\ketbra{01}+\ketbra{10})_{\textbf{X}_\xi \textbf{X}_{\xi'}}\otimes \mathds{1}_{J_X} \nonumber.
\end{align}
Here, a dilation step is in general required to write  the projectors $\Pi_X^{1_0}$ and  $\Pi_X^{1_1}$ before coarse-graining, the auxiliary system is the absorbed into one of the incoming junk systems, see appendix for details. With the help of Eqs.~(\ref{eq: rigid states},\ref{eq: rigid POVMS}) we express the output probabilities as 
\be\label{eq: cons projectors}
\begin{split}
P_Q(1_i,1_j,1_k) &=\frac{1}{8}\|\Pi^{1_i}_A \Pi^{1_j}_B\Pi^{1_k}_C(\ket{\Psi^c} +
\ket{\Psi^a})\|^2 \\ 
 P_Q(1_i,0,2)&=  \frac{1}{8}\|\Pi^{1_i}_X \ket{\Psi^a}\|^2 \quad\circlearrowright\\
P_Q(1_i,2,0) &=\frac{1}{8}\|\Pi^{1_i}_X \ket{\Psi^c}\|^ 2 \quad \circlearrowright
\end{split}
\ee
where we introduced the global states
\be
\begin{split}
\ket{\Psi^c} &\equiv\ket{01,01,01}_{\textbf{B}_\alpha \textbf{C}_\alpha \textbf{C}_\beta \textbf{A}_\beta  \textbf{A}_\gamma  \textbf{B}_\gamma}\ket{j_\alpha^c,j_\beta^c,j_\gamma^c}_{J_\alpha J_\beta J_\gamma}
\\\ket{\Psi^a} &\equiv\ket{10,10,10}_{\textbf{B}_\alpha \textbf{C}_\alpha \textbf{C}_\beta \textbf{A}_\beta  \textbf{A}_\gamma  \textbf{B}_\gamma}\ket{j_\alpha^a,j_\beta^a,j_\gamma^a}_{J_\alpha J_\beta J_\gamma}
\end{split}
\ee
corresponding to all the tokens sent clockwise (\textit{c}) or anticlockwise (\textit{a}). 
Here, the probabilities  
\be\begin{split}
    & 8 P_Q(1_i,1_j,1_k)= \|\Pi^{1_i}_A \Pi^{1_j}_B\Pi^{1_k}_C\ket{\Psi^c}\|^2\! \\ &+
\|\Pi^{1_i}_A \Pi^{1_j}_B\Pi^{1_k}_C\ket{\Psi^a}\|^2 +2\,  \text{Re}\bra{\Psi^c}\Pi^{1_i}_A \Pi^{1_j}_B\Pi^{1_k}_C \ket{\Psi^a} 
\label{eq: cons state 3} 
\end{split}\ee
are particularly interesting because they involve a coherence term between the global states $\ket{\Psi^a}$ and $\ket{\Psi^c}$, which only has a quantum interpretation. As $\Pi^{1_1}_X \ket{\Psi^a} =(\mathds{1}-\Pi^{1_0}_X-\Pi^{0}_X- \Pi^{2}_X) \ket{\Psi^a}= (\mathds{1}-\Pi^{1_0}_X)\ket{\Psi^a}$, and the states $\ket{\Psi^c}$ and $\ket{\Psi^a}$ are locally orthogonal on each party $\bra{\Psi^c} \Pi^x_X \Pi^y_Y\ket{\Psi^a}=0$, the coherence terms in Eq.~\eqref{eq: cons state 3} are equal up to a sign for all possible values $i,j$ and $k$. This allows us to quantify the coherence with a single value
\be\label{eq: coherence r}
r \equiv (-1)^{i+j+k} \,  2\, \text{Re}\bra{\Psi^c} \Pi^{1_i}_A \Pi^{1_j}_B\Pi^{1_k}_C \ket{\Psi^a}.
\ee

Remarkably, by adopting the nonlocality proof of~\cite{Renou_2019}  we derive a lower bound on this coherence
\be\label{eq: bound on r}
r \geq \frac{1}{2} \sin^3(\theta ) \big(3 \cos (\theta )+\cos (3 \theta )-6 \sin (\theta ) \big),
\ee
as a function of the parameter $\theta$, see Appendix D for full detail. The bound is the most stringent at $\theta_*\approx 0.36$, where $r\geq r_*$ with $r_* \approx 0.025$. The idea behind the derivation is to show that if $r$ is below the bound and Eqs.~\eqref{eq: cons projectors} hold, than $q_a(i,j,k)\equiv \bra{\Psi^a}\Pi^{1_i}_A \Pi^{1_j}_B\Pi^{1_k}_C  \ket{\Psi^a}$ and  $q_c(i,j,k)\equiv \bra{\Psi^c}\Pi^{1_i}_A \Pi^{1_j}_B\Pi^{1_k}_C \ket{\Psi^c}$ can not be valid probability distributions (not just for a triangle quantum model but in general). Since this last step of the argument ignores the network structure, it is not surprising that the bound~\eqref{eq: bound on r} we obtain is only nontrivial for the subset of distributions with $\theta\in(0,\theta_\text{max}\approx 0.48)$ -- the same subset where the nonlocality of the distribution has been proven in~\cite{Renou_2019}. The crucial difference is that it now applies to quantum models. Furthermore, by bounding the coherence $r$ we obtain a partial characterization of any quantum model underlying the RGB4 distribution. Quite an insightful one, as we will now see.\smallskip

\paragraph{Genuine network nonlocality --} Let us first show that the RGB4 distribution is GNNL, i.e. cannot be simulated by wiring of bipartite quantum boxes~\cite{Supic2022}. In fact, any such wiring results in measurements $\Pi^{x}_X$ that are separable for each party, e.g. $\Pi^{a}_A = \sum_k p_k \ketbra{\Psi_k^a}_{A_\beta}\otimes  \ketbra{\Phi_k^a}_{A_\gamma}$ for Alice. Since these measurements also satisfy the TC conditions~\eqref{eq: rigid POVMS}, it follows that $\braket{00}{\Psi_k^{1_i},\Phi_k^{1_i}}=\braket{11}{\Psi_k^{1_i},\Phi_k^{1_i}}=0$. Hence, these states are either of the form $\ket{\Psi_k^{1_i},\Phi_k^{1_i}}=\ket{01}_{\textbf{A}_\beta \textbf{A}_\gamma}\ket{\zeta}_{J_A}$ or $\ket{\Psi_k^{1_i},\Phi_k^{1_i}}=\ket{10}_{\textbf{A}_\beta \textbf{A}_\gamma}\ket{\zeta}_{J_A}$ for each $k$. But such measurements do not erase the information on the direction of each token, and give no coherence $\bra{\Psi^c}\Pi^{1_i}_A \Pi^{1_j}_B\Pi^{1_k}_C  \ket{\Psi^a} = 0$ (even if only one of the measurements $\Pi^{1_x}_X$ is separable). Hence, the distribution $P_Q(a,b,c)$ is genuinely network nonlocal if $r \neq 0$. \smallskip

\paragraph{Quantifying source entanglement --} Next we show that all the states distributed by the sources are entangled and quantify the amount of entanglement. To analyze the entanglement distributed by the sources we need a more precise description of the states. Let us decompose the junk system $J_\xi$ into some unknown auxiliary degrees of freedom $X'_\xi Y'_\xi$ that are indeed received and measured by the parties X and Y, and a system $E_\xi$ which can be controlled by an eavesdropper (Eve). Starting with
\be\nonumber
 \ket{\psi_\xi}= \frac{1}{\sqrt{2}}\big( \ket{01}_{\textbf{X}_\xi \textbf{Y}_\xi}\!\ket{j^{c}_\gamma}_{X'_\xi Y'_\xi E_\xi}\!\!+
\ket{10}_{\textbf{X}_\xi \textbf{Y}_\xi}\ket{j^{a}_\gamma}_{X'_\xi Y'_\xi E_\xi} \big)
\ee
and tracing out Eve's systems we define the states
\be
\rho^{(\xi)}_{\textbf{X}_\xi \textbf{Y}_\xi X'_\xi Y'_\xi}\equiv\tr_{E_\xi} \ketbra{\psi_\xi}
\ee
received by the parties. Knowing that the measurements act trivially on the system $E_\xi$ kept by the eavesdropper, we want to show that all these states are entangled. 

This can be shown by noting that if one state was separable the rigidity constraints would imply $r=0$. Instead, we will directly proceed to bound the  entanglement of the state $\rho^{(\alpha)}$ (or any of the other two) as quantified by its entanglement of formation $\cE_F$~\cite{EOF}. $\cE_F$ is an entanglement measure that for a mixed bipartite state $\rho_{BC}$ equals to the minimal average entropy of entanglement among all partitions of $\rho_{BC}$ in pure states, that is 
\be\begin{split}
&\cE_F(\rho_{B C}) \equiv \min \sum_k p_k S(\tr_B \ketbra{\psi_k})\\
&\text{such that }\quad     \rho_{BC}=\sum_k p_k \ketbra{\psi_k},
\end{split}
\ee
where $S$ is the von Neumann entropy. The rigidity constraint~\eqref{eq: rigid states} guarantees that each state in the partition of $\rho^{(\alpha)}$ is of the form $\ket{\psi_k}= \sqrt{q_k} \ket{01}_{\textbf{B}_\alpha\textbf{C}_\alpha} \ket{\phi_k}_{B'_\alpha C'_\alpha} +\sqrt{1-q_k} \ket{10}_{\textbf{B}_\alpha\textbf{C}_\alpha} \ket{\zeta_k}_{B'_\alpha C'_\alpha} $ for some unknown states $\ket{\phi_k}$ and $\ket{\zeta_k}$ of the auxiliary systems. Furthermore, the entropy of entanglement of this state is trivially bounded $S(\tr_{\textbf{B}_\alpha B'_\alpha} \ketbra{\psi_k})\geq h_\text{bin}(q_k)$ by the entropy $h_\text{bin}(q_k)$ of the binary probability distribution $(q_k,1-q_k)$. Hence the entanglement of formation satisfies
$\cE_F(\rho^{(\alpha)})\geq \min \sum_k p_k h_\text{bin}(q_k)$.
On top of that it is not difficult to see that the inequality~\eqref{eq: coherence r}, implies $\sum_{k} p_k \sqrt{q_k (1-q_k)} \geq 2 r$ for any partition of $\rho^{(\alpha)}$. It remains to minimize $\sum_k p_k h_\text{bin}(q_k)$ under the constraint $\sum_k p_k \sqrt{q_k(1-q_k)} \geq 2r$ to show that the entanglement of formation is lower bounded by
\be
\cE_F(\rho^{(\alpha)}) \geq h_\text{bin}\left(\frac{1}{2}(1-\sqrt{1-16 \, r^2})\right).
\ee
Hence, all sources must produce entanglement when $r\neq 0$. All the details of the derivation can be found in appendix E. For the maximal value $r_*$ certified by Eq.~\eqref{eq: bound on r}, we find that $\cE_F(\rho^{(\xi)}) > 2.5 \%$.

\paragraph{Quantifying output randomness --} Finally, let us bound the amount of randomness that is produced by the measurements. We focus on the entropy of a single output, say $a$. It is simpler to further coarse-grain the values of $a$ to define a bit $\bar a =0$ (for $a=0,2$) and $\bar a = 1$ (if $a=1_0,1_1$) encoded in the register $\bar A$, since Eq.~\eqref{eq: rigid POVMS} guarantees the junk degrees of freedom have no influence on $\bar a$. When tracing out all the systems but $\bar A E$ one finds a simple classical-quantum state
\begin{align}
\varrho_{\bar AE} &= \frac{1}{2} \ketbra{0}_{\bar A}\otimes \rho_{E|\bar a=0}  + \frac{1}{2} \ketbra{1}_{\bar A} \otimes \rho_{E|\bar a=1} \\
\rho_{E|\bar a=0} &= \frac{1}{2}(\rho_{E_{\beta\gamma}}^{ca}+\rho_{E_{\beta\gamma}}^{ac}), \, \nonumber 
\rho_{E|\bar a=1} =
 \frac{1}{2}(\rho_{E_{\beta\gamma}}^{cc}+\rho_{E_{\beta\gamma}}^{aa})
\end{align}
where $\rho_{E_{\beta\gamma}}^{xy} =  \rho_{E_\beta}^x \otimes  \rho_{E_\gamma}^y$ with  $\rho_{E_\xi}^x = \tr_{X'_\xi Y_\xi'} \ketbra{j^x_\xi}_{X
 _\xi Y_\xi' E_\xi}$. Eve's conditional min-entropy~\cite{Konig2009} is related by
$H_\text{min}(\bar A|E) = -\log_2\left(\frac{1}{2}(1+D(\rho_{E|\bar a=0},\rho_{E|\bar a=1}))\right) $
to the trace distance $D$ between her marginal states. Clearly, the entropy is not zero, as Eve's perfect knowledge of the direction of tokens ($D= 1$) would imply no coherence ($r=0$). Nevertheless, we found that the technical challenge of deriving a decent upper bound on $D$ from a lower bound on $r$ is not straightforward. In appendix E we show that
$D(\rho_{E|\bar a=0},\rho_{E|\bar a=1}) \geq\sqrt{1-4 r}$, leading to 
\be
H_\text{min}(\bar A|E) \geq -\log_2\left(\frac{1}{2}(1+\sqrt{1-4r}) \right).
\ee
For the maximal value $r_*$ we find $H_\text{min}(\bar A|E)\geq 3.8\%$.

\paragraph{Generalizations --}
The above partial self-testing results can be generalized to any ring network. The proofs of Results 1 and 2 are given in Appendix B and C, respectively, with notations given in Appendix A.

Another generalization concerns Parity Token Counting (PTC) distributions on the triangle~\cite{boreiri2022}, for which the equivalent of Result 2 also holds and is particularly simple to prove; see Appendix F. In a PTC strategy each source has a single token, and the parties only output the parity of the total number of received tokens.

We expect these results to be helpful to characterize various quantum distributions that become (P)TC upon coarse-graining, similarly to our analysis of RGB4.

\paragraph{Conclusion and Outlook --}
We showed that quantum nonlocal distributions on ring networks without inputs can be partially self-tested. Applying these methods to the triangle network, we prove that the nonlocal distribution of RGB4 (from Ref. \cite{Renou_2019}) has interesting properties. First, all measurements must be entangled, hence demonstrating GNNL. Also, all states must be entangled, with a lower bound on their entanglement. Finally, we obtain a lower bound on the min-entropy for a local outcome, hence quantifying the amount of randomness.

All the above results can in principle be strengthened quantitatively by obtaining tighter bounds on the parameter $r$. This could be done by better exploiting the triangle structure, or by considering other nonlocal variants of the distribution \cite{Pozas2022}. 

Another interesting question is whether the RGB4 can be proven to be FNNL. Here we show a first step in this direction, namely that if the experiment abides by quantum physics then all sources must produce entanglement. But can one prove that all sources must be nonlocal, even if stronger-than-quantum non-signaling resources are accessible? A related question is to show that the RGB4 distribution is genuine network nonlocal when considering sources that produce non-signaling correlations and local wirings \cite{Supic2022}.

Finally, it would be desirable to 
make our results robust to noise. A first step could be to obtain approximate rigidity results for (P)TC.

\emph{Acknowledgements.---} We thank Marc-Olivier Renou  for discussions and Victor Gitton for helpful comments on the first version of the manuscript. We acknowledge financial support from the Swiss National Science Foundation (project 2000021\_192244/1 and NCCR SwissMAP).

\bibliography{qnets}
\begin{widetext}

\section{Appendix A. Ring networks}
\label{app: rings}
A ring network depicted in Fig.~\ref{fig:ring} is a straightforward generalization of the triangle. For a $n$-partite ring, each party $k\in\{1,\dots, n\}$ receives the systems $L_k$ and $R_k$ form the neighbouring sources and outputs $a_k$. Each source $S_k$ prepares the state of the systems $R_k$ and $L_{k+1}$. For the labels of the systems, the addition is meant modulo $n$, e.g. $L_{n+1}$ and $L_1$ are the same systems.

\begin{figure}[h]
    \centering
    \includegraphics[width=0.45\columnwidth]{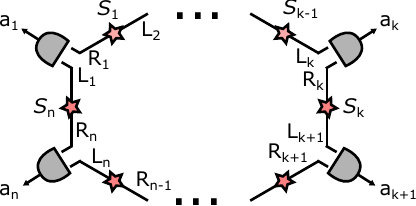}
    \caption{A ring network. Each source $S_k$ prepares a state of the systems $R_k$ and $L_{k+1}$; each party measures the systems $L_k R_k$ to produce the output $a_k$. In the case of quantum models, the bipartite state produced by each source $S_k$ is denoted $\ket{\psi_k}$, and the global state of all the systems before the measurements is called $\ket{\Psi}$.}
    \label{fig:ring}
\end{figure}

\subsection{Token Counting strategies }  The classical TC strategies we mentioned for the triangle can be straightforwardly generalized to the ring. Each source $S_k$ has $N_k$ tokens. With probability $p_k(t)$ it sends $i$ tokens to $R_k$ and $(N_k-t)$ tokens to $L_{k+1}$. Each party outputs the total number of received tokens. By construction, one has
\be
a_1+\dots +a_n = N \equiv N_1+\dots + N_n.
\ee

\subsection{A priori quantum description of the setup}

By assumption in the scenario we consider each source $S_k$ controls two finite dimensional quantum systems $R_k$ and $L_{k+1}$. It prepares them in a state which a priori can not be assumed pure. Denote it $\varrho_k$ -- a density operator on the tensor product of the finite dimensional Hilbert spaces associated to the systems. By introducing an auxiliary finite-dimensional system $A_k$ the  state $\varrho_k$ can be purified to $\ket{\psi_k}$, i.e. $\varrho_k = \tr_{A_k} \ketbra{\psi_k}_{R_k L_{k+1}A_k}$.  Furthermore, without loss of generality the auxiliary system $A_k$ can be absorbed in either $R_k$ or $L_{k+1}$. The dimension of the Hilbert space that describes it is them augmented, but remains finite. In other words, in the device-independent setting we consider, the state 
\be
\ket{\psi_k}_{R_k L_{k+1}} \in \cH_{R_k}\otimes \cH_{L_{k+1}}
\ee 
prepared by the source $S_k$ can be assumed pure from the beginning. Here we also introduced the Hilbert spaces $ \cH_{R_k}$ and $ \cH_{L_{k+1}}$ associated to each system.

Let us now consider the description of a measurements performed by the $k$-th party on the system $X_k=L_k R_k$ to produce the output $a_k$. A priori they are given by a positive operator valued measure (POVM), that is a set of non-negative hermitian operators $E^{a}_{X_k}: \cH_{L_k}\otimes \cH_{R_k} \to \cH_{L_k}\otimes \cH_{R_k}$
\be
\{E_{X_k}^a\}_{a}, \text{ such that } \qquad 
E_{X_k}^a={E_{X_k}^a}^\dag \succeq 0  \qquad \sum_{a} E_{X_k}^a = \mathds{1}_{X_k}
\ee
 By Steinspring dilation theorem this POVM can be dilated to a projective measurement. This requires the introduction of an auxiliary system $M_k$ prepared in the state $\ket{0}$, with associated finite dimensional Hilbert space $\cH_{M_k}$. For any state $\rho_{L_k R_k}$  the probability of an outcome $a_k$ is then given by 
\be
 E_{X_k}^a  = \tr_{M_k} \bar \Pi_{X_k}^a \ketbra{0}_{M_k}.
\ee
Here $\{\bar \Pi_{a_k}\}_{a_k}$ is a projector valued measure (PVM), i.e. in addition to being a POVM it fulfills $(\bar \Pi_{X_k}^a)^2 = \bar \Pi_{X_k}^a$, with the projector acting on the systems $L_k R_k M_k$.
Below we will also deal with unitaries $\bar U_k$ acting on the same systems 
\be
\bar U_k= \sum_x e^{\ii \varphi_x} \bar \Pi_{X_k}^x,
\ee
here it is only important that the phases $e^{\ii \varphi_x}$ are different for different $x$.

\subsection{Proof of RESULT 0}

\textbf{Result 0} \textit{For a network state $\ket{\Psi}=\ket{\psi_1}\dots\ket{\psi_n}$ and unitaries $\bar U_k = \sum_{x} e^{\ii \varphi_x} \bar \Pi_{X_k}^x$ with $e^{\ii \varphi_x} \neq e^{\ii \varphi_{x'}}$ for $x\neq x'$ the identity
\be\label{eq: rigid app 1}
\left(\bigotimes_{k=1}^n \bar U_{k} \right) \ket{\Psi}\bigotimes_k\ket{0}_{M_k} = \ket{\Psi}\bigotimes_k\ket{0}_{M_k}
\ee
implies that the original measurements are projective}
\be
E_{X}^x = \bra{0}_{M_X} \bar \Pi_X^x \ket{0}_{M_X} = \Pi_{X}^x.
\ee
Consider the action of a single unitary $\bar U_k$ on the states $\ket{\psi_{k-1}}_{R_{k-1}L_k}\ket{\psi_k}_{R_k L_{k+1}} \ket{0}_{M_k}$. It is the only unitary acting on the system $M_k$ and by virtue of Eq.~\eqref{eq: rigid app 1} we know that it leaves it in the state $\ket{0}_{M_k}$. Formally, this can be expressed as
\be
\bar U_k \ket{\Psi}\bigotimes_i \ket{0}_{M_i} = \left(\bigotimes_{j\neq k} \bar U_j^\dag\right) \ket{\Psi}\bigotimes_i \ket{0}_{M_i} = \ket{0}_{M_k}\left(\bigotimes_{j\neq k} \bar U_j^\dag \ket{\Psi}\bigotimes_{i\neq k} \ket{0}_{M_i}\right)
\ee
and therefore
\be\label{eq: app result 0 1}
\left(\mathds{1}_{R_{k-1}L_{k+1}}\otimes \bar U_k \right)\ket{\psi_{k-1}}_{R_{k-1}L_k}\ket{\psi_k}_{R_k L_{k+1}} \ket{0}_{M_k}=\ket{\xi'}_{R_{k-1}L_kR_k L_{k+1}}\ket{0}_{M_k}.
\ee
for some state $\ket{\xi'}_{R_{k-1}L_kR_k L_{k+1}}$.  Since the density operators $\tr_{R_{k-1}} \ketbra{\psi_{k-1}}_{R_{k-1}L_k}$ and $\tr_{L_{k+1}} \ketbra{\psi_{k}}_{R_{k}L_{k+1}}$ have full support on the Hilbert spaces $\cH_{L_k}$ and $\cH_{R_k}$ (by definition of these Hilbert spaces), Eq.~\eqref{eq: app result 0 1} implies that for any state $\ket{\ell}_{L_k}\in \cH_{L_k}$ and $\ket{r}_{R_k} \in \cH_{R_k}$ 
\be
\bar U_k \ket{\ell}_{L_k} \ket{r}_{R_k}\ket{0}_{M_k}= \ket{\xi_{\ell,r}}_{L_k R_k}\ket{0}_{M_k},
\ee
with some state $\ket{\xi_{\ell,r}}_{L_k R_k}\in \cH_{L_k}\otimes \cH_{R_k}$. It is the easiest to see with state $\ket{\ell}_{L_k}$ and $\ket{r}_{R_k}$ that appear in the Schmidt decomposition of $\ket{\psi_{k-1}}_{R_{k-1}L_k}$ and $\ket{\psi_{k}}_{R_{k}L_{k+1}}$ respectively.

Starting with two orthonormal bases $\{\ket{\ell}_{L_k}\}_\ell$ of $\cH_{L_k}$ and $\{\ket{r}_{R_k}\}_r$ of $\cH_{R_k}$ we see that the states $\{ \ket{\xi_{r,\ell}}\}_{r,\ell} $ form an orthonormal basis of $\cH_{R_k}\otimes \cH_{L_k}$
\be
\delta_{r,r'}\delta_{\ell,\ell'}= \braket{\ell, r,0}{\ell',r',0}_{L_k,R_k,M_k}= \bra{\ell, r,0} \bar U_k^\dag \bar U_k\ket{\ell',r',0} = \braket{\xi_{r,\ell}}{\xi_{r',\ell'}}
\ee
This guarantees that the operator $U_k$ defined as the restriction
\be\begin{split}
 U_k &= \bra{0}_{M_k} \bar U_k \ket{0}_{M_k} = \sum_{\ell,r} \ketbra{\xi_{\ell,r}}{\ell,r}\\
U_k &: \cH_{L_k}\otimes \cH_{R_k} \to \cH_{L_k}\otimes \cH_{R_k} 
\end{split}\ee
is also unitary (a basis change). By construction this operator is also equal to 
\be
U_X = \bra{0}_{M_k} \bar U_k \ket{0}_{M_k} = \sum_x e^{\ii \varphi_x} \bra{0}_{M_k} \bar \Pi_{X_k}^x \ket{0}_{M_k} = \sum_x e^{\ii \varphi_x} E_{X_k}^x.
\ee
The following lemma shows that an operator $U_X =\sum_x e^{\ii \varphi_x} E_{X_k}^x$ is unitary if and only if $\{E_{X_k}^x\}_x$ is a PVM. Which proves the result $\square$.\bigskip

\textbf{Lemma 0} \textit{Consider a POVM $\{ E_x\}_x$ and the operator 
$U = \sum_x e^{\ii \varphi_x} E_x $
with all $e^{\ii \varphi_x}$ different for different $x$. $U$ is unitary if and only if $\{ E_x= \Pi_x\}_x$ is a PVM.}\\

Since $U$ is unitary it can be diagonalized. Hence, there is a basis $\{\ket{k}\}_k$ of the underlying Hilbert space such that 
\be
U \ket{k} = e^{\ii \lambda_k}\ket{k}
\ee
for some real  $\lambda_k\in[0,2\pi)$. For all $k$ we thus have
\be
e^{\ii \lambda_k} = \bra{k} U \ket{k} =  \bra{k} \sum_x e^{\ii \varphi_x} E_x \ket{k} = \sum_x e^{\ii \varphi_x} p_k(x),
\ee
where $p_k(x) =\bra{k} E_x \ket{k}$ is a probability distribution. In particular, it implies 
\be
1 = |e^{\ii \lambda_k}| = |\sum_x e^{\ii \varphi_x} p_k(x)|,
\ee
since all $e^{\ii \varphi_x}$ are different and $\sum_x p_k(x) = 1$, this equality is only possible if $p_k(x) = \delta_{x,y(k)}$  i.e. nonzero for only a single $x = y(k)$. Therefore, $\|\sqrt{E_x}\ket{k}\|=\delta_{x,y(k)}$ and
\be
E_x \ket{k} = \begin{cases}
\ket{k} & x =y(k) \\
0 & x \neq y(k).
\end{cases}
\ee
Hence the operators $E_x$ orthogonal projector  diagonal in the basis $\{\ket{k}\}_k$. This shows the "only if" direction. The converse is trivial as any $\sum_x e^{\ii \varphi_x} \Pi_x$ is unitary. $\square$

\section{Appendix B. Proof of result 1.}
\label{app: theorem 1}

\textbf{Result 1.} \textit{Consider a quantum state $\ket{\Psi}=\ket{\psi_1}\dots\ket{\psi_n}$ of a $n$-partite ring network, where each party receives two systems $L_k$ and $R_k$ from two different sources, and a collection of unitary operators $U_{1},\dots, U_{n}$ applied by each party.  The  condition}
\be\label{eq: rigid app}
\bigotimes_{k=1}^n U_{k} \ket{\Psi} = \ket{\Psi}
\ee
\textit{implies that all the unitaries are products}
\be
U_{k} = V_{L_k}^{(k)}\otimes W_{R_k}^{(k)}.
\ee

We  prove this statement in this section. To start we will introduce some notation and useful operators. We put each state $\ket{\psi_k}$ in the Schmidt diagonal form 
\be
\ket{\psi_k}= \sum_{i=1}^{d_k} \lambda_i^{(k)}\ket{i,i}_{R_k L_{k+1}}
\ee
for some finite $d_k$ and $\lambda_i^{(k)}>0$, it defines a basis for each of the systems. It also allows us to rewrite the states as 
\be
\ket{\psi_k}=\Lambda^{(k)}\otimes \mathds{1} \ket{\Omega} =\mathds{1}\otimes \Lambda^{(k)} \ket{\omega_k},
\ee
where $\ket{\omega_k} = \frac{1}{\sqrt{d_k}}\sum_{i=1}^{d_k} \ket{i,i}_{R_k L_{k+1}}$ and $\Lambda^{(k)}$ is diagonal in the Schmidt basis $\Lambda^{(k)}\ket{i}_{R_k} = \sqrt{d_k}\lambda_{k}^{(i)}\ket{i}_{R_k}$ 
is diagonal in the
Schmidt basis and invertible. With the help of the identity $\mathds{1}\otimes M \ket{\omega} =M^T\otimes\mathds{1} \ket{\omega}$ for a maximally entangled state $\ket{\omega}$ and the transpose with respect to the Schmidt basis, we can now move the unitaries to act on different systems. Concretely, we can write
\be\label{def: G_k}\begin{split}
(U_k)_{L_k R_k}\ket{\psi_{k-1}}_{R_{k-1}L_k}\ket{\psi_k}_{R_kL_{k+1}} 
& =\Lambda^{(k-1)}_{R_{k-1}} \otimes (U_k)_{L_k R_k}\otimes \Lambda^{(k)}_{L_{k+1}}\ket{\omega_{k-1}}\ket{\omega_k}\\
& = (\Lambda^{(k-1)}_{R_{k-1}}\otimes \Lambda^{(k)}_{L_{k+1}})  (U_k^T)_{R_{k-1} L_{k+1}} \ket{\omega_{k-1}}\ket{\omega_k}\\
&= (G_k)_{R_{k-1}L_{k+1}}\ket{\omega_{k-1}}\ket{\omega_k}
\end{split}
\ee
where in the last line we defined the invertible operator 
\be
G_k \equiv (\Lambda^{(k-1)}\otimes \Lambda^{(k)}) U_k^T
\ee
acting on the systems $R_{k-1} L_{k+1}$. Similarly, we rewrite
\be\label{def: F_k}\begin{split}
 (U_k)^\dag_{L_k R_k} \ket{\psi_{k-1}}_{R_{k-1}L_k}\ket{\psi_k}_{R_kL_{k+1}} 
& = (U_k)^\dag_{L_k R_k}( \Lambda^{(k-1)}_{L_k}\otimes \Lambda^{(k)}_{R_k}) \ket{\omega_{k-1}}\ket{\omega_k}\\
&= (F_k)_{L_{k}R_{k}}\ket{\omega_{k-1}}\ket{\omega_k}
\end{split}
\ee
where we introduced the invertible operator 
\be
F_k \equiv U^\dag_k( \Lambda^{(k-1)}\otimes \Lambda^{(k)})
\ee
acting on the systems $L_kR_k$.  For the following, it is convenient to introduce the notation
\be
\ket{\Omega}\equiv \ket{\omega_1}\dots\ket{\omega_n}.
\ee
Next, we will consider even and odd rings separately in the next two sections. The reasoning leading to the proof is  graphically summarized in Fig.~\ref{fig:ring proofs}, which can be a useful guide to the reader.
In both cases (even and odd) to make the final step, we invoke one of the two technical lemmas on the rigidity of a chain of operators acting on several systems. The lemmas will be stated at the moment where we use them, while their proofs are given at the very end of this appendix. 

\begin{figure*}[t!]
    \centering
    \includegraphics[width=0.9 \textwidth]{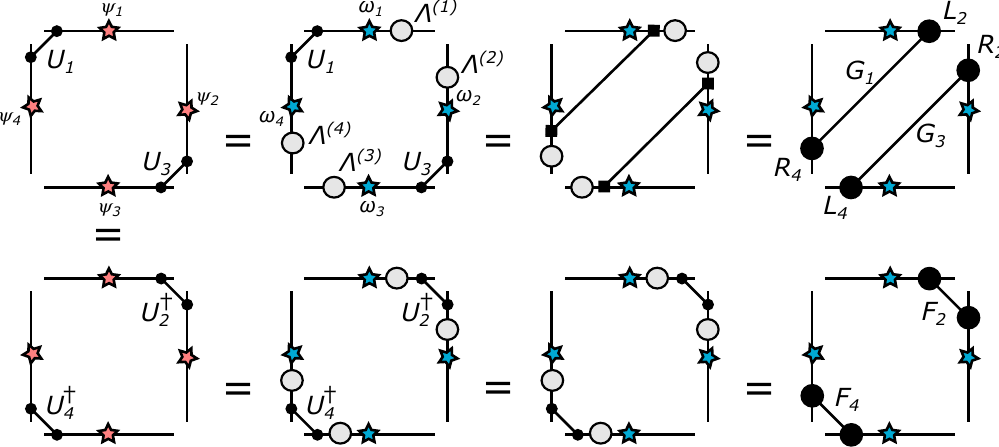}\\
    \vspace{10 mm}
    \includegraphics[width= 0.9 \textwidth]{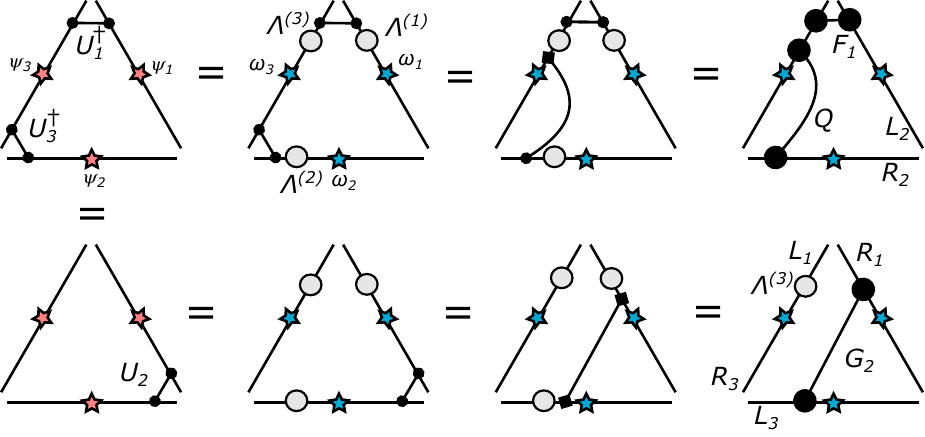}
    \caption{(Square) Even rings -- minimal example of the square network: A graphical summary of transformations leading from the equation $U_1U_3\ket{\Psi}=U_2^\dag U_4^\dag\ket{\Psi}$~\eqref{eq: app even} -- the left column -- to the equation $(F_1)_{R_4 L_2} (F_3)_{R_2 L_4}\ket{\Omega}=(G_2)_{L_2 R_2} (G_4)_{L_4R_4}\ket{\Omega}$~(\ref{eq: app even even},\ref{eq: app even odd}) -- the right column.
   (Triangle) Odd rings -- minimal example of the triangle network: A graphical summary of transformations leading from the equation $U_1^\dag U_3^\dag \ket{\Psi}= U_2\ket{\Psi}$~\eqref{eq: odd} -- the left column -- to the equation $(F_1)_{L_1R_1}Q_{L_1L_3}\ket{\Omega} = (G_2)_{R_1L_3} \ket{\Omega}$~\eqref{eq: app odd final} -- the right column.
   (Both)  $\ket{\omega_k}$ are maximally entangled states and $\Lambda^{(k)}$ are invertible local operators such that $\mathds{1}\otimes \Lambda^{(k)}\ket{\omega_k} =  \Lambda^{(k)}\otimes \mathds{1}\ket{\omega_k} = \ket{\psi_k}$. We use the property $\mathds{1}\otimes M \ket{\omega}= M^T \otimes\mathds{1} \ket{\omega}$ to change the systems on which the unitaries act.}
    \label{fig:ring proofs}
\end{figure*}

\subsection{Even rings}

For even rings $(n = 2m, m\geq 2)$, by multiplying both side of the equation~\eqref{eq: rigid app} by $U_2^\dag U_4^\dag\dots U_{2m}^\dag$  we rewrite it as
\be \label{eq: app even}
U_1 U_3 \dots U_{2m-1}\ket{\Psi}= U_2^\dag U_4^\dag\dots U_{2m}^\dag \ket{\Psi}.
\ee

Here, the unitaries only act on odd parties (lhs) or even parties (rhs).
Now, using the definitions (\ref{def: G_k},\ref{def: F_k}) we rewrite both sides of the equation as
\begin{align} 
U_1 U_3 \dots U_{2m-1}\ket{\Psi} 
&=  (G_1)_{R_{2m} L_2}(G_3)_{R_2 L_4}\dots (G_{2m-1})_{R_{2m-2}L_{2m}}
\ket{\Omega}, \label{eq: app even even}\\
  U_2^\dag U_4^\dag\dots U_{2m}^\dag \ket{\Psi}
&= (F_2)_{L_2R_2}(F_4)_{L_4R_4}\dots (F_{2m})_{L_{2m}R_{2m}}\ket{ \Omega}. \label{eq: app even odd}
\end{align}
Note that on the right hand side of these equations the operators only act on the even parties $2k$. Using Choi–Jamio\l{}kowski duality, $\mathds{1}\otimes A \ket{\omega} = \mathds{1}\otimes B \ket{\omega} \implies A=B$, we combine Eqs.~(\ref{eq: app even},\ref{eq: app even even},\ref{eq: app even odd}) to obtain the following equality between the chains of operators
\be\label{eq: app even final}
(F_2)_{L_2R_2}\otimes(F_4)_{L_4R_4}\dots \otimes(F_{2m})_{L_{2m}R_{2m}} =    
(G_1)_{R_{2m} L_2}\otimes(G_3)_{R_2 L_4}\dots \otimes(G_{2m-1})_{R_{2m-2}L_{2m}}. 
\ee
In order to continue, we need the following Lemma.

{\bf Lemma 1e} {\it Consider $n=2m$ quantum systems and two sets $m$ nonzero bipartite  operators $A_{1,2}, A_{3,4}, \dots, A_{2m-1, 2m}$ and $B_{2,3}, B_{4,5}, \dots, B_{2m, 1}$ where the labels indicate on which pair of systems the operators act. If the following identity is satisfied 
\be\label{eq: conition even}
\bigotimes_{k=1}^m A_{2k-1,2k} = 
\bigotimes_{k=1}^m B_{2k,2k+1},
\ee
then all the operators are product}
\be\begin{split}
A_{2k-1,2k}&= \text{A}_{2k-1}\otimes \text{A}'_{2k}\\
B_{2k,2k+1}&= \text{B}_{2k}\otimes \text{B}'_{2k+1}
\end{split}
\ee
A graphical illustration of the Lemma is given in Fig.~\ref{fig:ring lemmas}(a), while its proof can be found in Sec.~\ref{sec: proof 1e}.

To apply Lemma 1e, note that in Eq.~\eqref{eq: app even final}
we are precisely dealing with $n=2m$ quantum systems, that can be arranged as $L_2,R_2,L_4,R_4,\dots, L_{2m},R_{2m}$ and two collections of operators $F_2,\dots F_{2m}$ (playing the roles of $A_{2k-1,2k}$) and $G_3,\dots, G_{2m-1},G_1$ (playing the roles of $B_{2k,2k+1}$) acing on neighbouring pairs of systems. Hence, we conclude that $F_k$ and $G_k$ are products, i.e.
\be
\begin{split}
(G_{2k+1})_{R_{2k} L_{2k+2}} &= (\text{G}_{2k+1})_{L_{2k}} \otimes (\text{G}'_{2k+1})_{R_{2k+2}}\\
   (F_{2k})_{L_{2k}R_{2k}} &= (\text{F}_{2k})_{L_{2k}} \otimes (\text{G}'_{2k})_{R_{2k}}.
\end{split}
\ee
From the definition of $G_k$ and $F_k$ it is straightforward to see that all the $U_k$ are then also products (recall that all the operators are invertible). This proves result 1 for even rings.

\subsection{Odd rings}

Let us now consider odd rings ($n=2m+1$, $n\geq 3$), which includes the triangle network. We rewrite the equation~\eqref{eq: rigid app} as
\be\label{eq: odd}
U_2 U_4 \dots U_{2m} \ket{\Psi} =
U_1^\dag U_3^\dag\dots U_{2m+1}^\dag\ket{\Psi}.
\ee
Similarly, to the even ring case, the left hand side can be rewritten as
\be
U_2 U_4 \dots U_{2m} \ket{\Psi} 
= 
(G_2)_{R_1L_3}(G_4)_{R_3L_5}\dots (G_{2m})_{R_{2m-1}L_{2m+1}} \Lambda^{(2m+1)}_{R_{2m+1}}\ket{\Omega},
\ee
using  Eq.~\eqref{def: G_k}. 

A notable difference with the case of even rings, is that the right hand side of Eq.~\eqref{eq: odd} involves the operators $U_1^\dag$ and $U_{2m+1}^\dag$ that act of the same state $\ket{\psi_{2m+1}}$. In order to bring it to the desired form we proceed in steps. First rewrite
\be\label{eq: app odd temp}
 U_3^\dag\dots U_{2m-1}^\dag\ket{\Psi} = (F_3)_{L_3 R_3}\dots (F_{2m-1})_{R_{2m-1}L_{2m-1}} \ket{\Omega},
\ee
where the operators $(F_3)_{L_3 R_3}\dots (F_{2m-1})_{R_{2m-1}L_{2m-1}}$
act trivially on the systems $R_{2m}L_{2m+1}, R_{2m+1}L_1$ and $R_1 L_2$ prepared by the sources $S_{2m},S_{2m+1}$ and $S_1$. 

For the unitaries $U_1 ^\dag$ and $U_{2m+1}^\dag$ acting on these systems we have
\be\begin{split}
U_1^\dag U_{2m+1}^\dag \ket{\psi_{2m}}_{R_{2m}L_{2m+1}}\ket{\psi_{2m+1}}_{R_{2m+1}L_{1}}\ket{\psi_1}_{R_1L_2} 
&=  U_1^\dag U_{2m+1}^\dag (\Lambda^{(2m)}_{L_{2m+1}}\otimes\Lambda^{(2m+1)}_{L_1}\otimes \Lambda^{(1)}_{R_1}) \ket{\omega_{2m}}\ket{\omega_{2m+1}}\ket{\omega_1} \\
&= (U_1^\dag \Lambda^{(2m+1)}_{L_1}\otimes \Lambda^{(1)}_{R_1})(U_{2m+1}^\dag \Lambda^{(2m)}_{L_{2m+1}}) \ket{\omega_{2m}}\ket{\omega_{2m+1}}\ket{\omega_1} \\
& =(F_1)_{L_1 R_1}(U_{2m+1}^\dag \Lambda^{(2m)}_{L_{2m+1}}) \ket{\omega_{2m}}\ket{\omega_{2m+1}}\ket{\omega_1}
\end{split}
\ee
We want to rewrite this expression in such a way that nothing acts on the system $R_{2m+1}$. This can be done by defining
\be
(U_{2m+1}^\dag)_{L_{2m+1}R_{2m+1}} \Lambda^{(2m)}_{L_{2m+1}} \ket{\omega_{2m}}_{R_{2m}L_{2m+1}}\ket{\omega_{2m+1}}_{R_{2m+1}L_1}   
=
Q_{L_{2m+1}L_1} \ket{\omega_{2m+1}}_{R_{2m+1}L_1}\ket{\omega_1}_{R_1L_2}
\ee
with $Q_{AB} \equiv ((U_{2m+1}^\dag)_{AB}\Lambda^{(2m)}_{A})^{T_B}$ is the partially transposed operator (with respect to the system $B$, in our case $R_{2m+1}$). $Q_{L_{2m+1}L_1}$ is a product of invertible operators, and is thus invertible. With the help of this operator we write
\be
U_1^\dag U_{2m+1}^\dag \ket{\psi_{2m}}_{R_{2m}L_{2m+1}}\ket{\psi_{2m+1}}_{R_{2m+1}L_{1}}\ket{\psi_1}_{R_1L_2} 
= (F_{1})_{L_1R_1} Q_{L_{2m+1}L_1} \ket{\omega_{2m}}\ket{\omega_{2m+1}}\ket{\omega_1}.
\ee
Finally, combing this realtion with Eq.~\eqref{eq: app odd temp} in order to obtain
\be
U_1^\dag U_3^\dag\dots U_{2m+1}^\dag\ket{\Psi} 
= \left(\bigotimes_{k=1}^m ( F_{2k-1})_{L_{2k-1}R_{2k-1}}\right) Q_{L_{2m+1}L_1}\ket{\Omega}.  
\ee

Coming back to the condition~\eqref{eq: odd} we get the equality,
\be
   (G_2)_{R_1L_3}(G_4)_{R_3L_5}\dots (G_{2m})_{R_{2m-1}L_{2m+1}} \Lambda^{(2m+1)}_{R_{2m+1}} \ket{\Omega}
   = \left(\bigotimes_{k=1}^m ( F_{2k-1})_{L_{2k-1}R_{2k-1}}\right) Q_{R_{2m+1}R_1}\ket{\Omega}
\ee
on both side of the equation all the operators act trivially on the systems $L_{2k}$, $R_{2k}$, and $L_1$, prepared by the sources $S_{2k-1},S_{2k}$ (for $k=1,\dots,m$) and $S_{2m+1}$ respectively. In other words, for all the involved bipartite maximally entangled states $\ket{\omega_k}$ the operators only act on one of the two systems. By Choi–Jamio\l{}kowski duality we thus arrive at the operator identity
\be
   (G_2)_{R_1L_3}(G_4)_{R_3L_5}\dots (G_{2m})_{R_{2m-1}L_{2m+1}} \Lambda^{(2m+1)}_{R_{2m+1}} 
   = \left(\bigotimes_{k=1}^m ( F_{2k-1})_{L_{2k-1}R_{2k-1}}\right) Q_{R_{2m+1}R_1}
\ee
 Next, we  multiply both sides by $\left(\Lambda^{(2m+1)}_{R_{2m+1}} \right)^{-1}$, and define 
\be
\widetilde{Q}_{R_{2m+1}R_1}\equiv \left(\Lambda^{(2m+1)}_{R_{2m+1}} \right)^{-1} Q_{R_{2m+1}R_1}
\ee
to write
\be\label{eq: app odd final}
 \left(\bigotimes_{k=1}^m ( F_{2k-1})_{L_{2k-1}R_{2k-1}}\right) \widetilde{Q}_{R_{2m+1}R_1} = (G_2)_{R_1L_3}(G_4)_{R_3L_5}\dots (G_{2m})_{R_{2m-1}L_{2m+1}}.
\ee
To continue the reasoning we use the following lemma.
\smallskip

{\bf Lemma 1o} {\it Consider $n=2m+1$ quantum systems, a set of $m+1$ nonzero bipartite  operators $A_{1,2}, A_{3,4}, \dots, A_{2m-1,2m}, A_{2m+1, 1}$ and a set of $m$ nonzero operators $B_{2,3}, B_{4,5}, \dots, B_{2m,2m+1}$ where the labels indicate on which pair of systems the operators act. If the following identity is satisfied 
\be\label{eq: conition odd}
\left(\bigotimes_{k=1}^m A_{2k-1,2k} \right) A_{2m+1,1} = 
\bigotimes_{k=1}^m B_{2k,2k+1},
\ee
then all the operators are product}
\be\begin{split}
A_{2k-1,2k}&= \text{A}_{2k-1}\otimes \text{A}'_{2k}\\
B_{2k,2k+1}&= \text{B}_{2k}\otimes \text{B}_{2k+1}'
\end{split}
\ee
The proof of the lemma is postponed to the end of the current appendix, while an illustration is given in Fig.~\ref{fig:ring lemmas}(b).
\smallskip

To apply Lemma 1o to our situation, given bty Eq.~\eqref{eq: app odd final}, one identifies the $n=2m+1$ quantum system of the Lemma with $R_1, L_3,R_3,\dots,L_{2m-1},R_{2m-1}$.
Then the bipartite operators $F_1,\dots, F_{2m-1}, \widetilde{Q}$ play the role of $A_{1,2},\dots,A_{2m-1,2m}, A_{2m+1,1}$, and $G_2,\dots, G_{2m}$ play the role of $B_{2,3},\dots B_{2m,2m+1}$. Lemma 1o guarantees that all of these operators are products. Finally, it is straightforward to conclude that all of the unitaries $U_1,
\dots, U_{2m+1}$ are also products for odd rings. This concludes the proof of Result 1 $\square$.

\begin{figure}
    \centering
    \includegraphics[width=0.5\columnwidth]{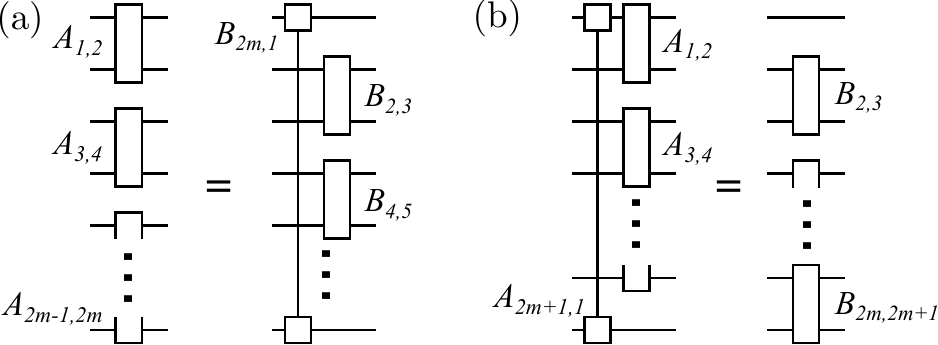}
\caption{Illustration of the Lemmas 1e and 1o for even (a) and odd (b) number of parties. Given any two collections of nonzero bipartite operators $\{A_{1,2},\dots, A_{2k-1,2k}\dots\}$ and $\{B_{2,3},\dots, B_{2k,2k+1},\dots\}$, if the operator equalities~(\ref{eq: conition even},\ref{eq: conition odd}) illustrated in the figure are satisfied, the lemmas guarantee that all the bipartite operators are product $A_{2k-1,2k}= \text{A}_{2k-1}\otimes \text{A}'_{2k}$
and $B_{2k,2k+1}= \text{B}_{2k}\otimes \text{B}'_{2k+1}$ }
    \label{fig:ring lemmas}
\end{figure}
\bigskip

\subsection{Proof of Lemma 1e}
\label{sec: proof 1e}
 Consider $n=2m$ quantum systems and two sets $m$ of nonzero bipartite operators $A_{1,2}, A_{3,4}, \dots, A_{2m-1, 2m}$ and $B_{2,3}, B_{4,5}, \dots, B_{2m, 1}$ which satisfy 
\be
\bigotimes_{k=1}^m A_{2k-1,2k} = 
\bigotimes_{k=1}^m B_{2k,2k+1},
\ee
where the labels indicate on which pair of systems the operators act.

To show that the operators are product introduce an orthonormal operator basis $\{X_{2k}^{(i)}\}_{i}$ (such that $\tr X^{(i)}_{2k} X^{(j)\dag}_{2k}=\delta_{ij}$) for each even system. It allows us to decompose each operator as
\be\label{eq app: init A B}\begin{split} 
    A_{2k-1,2k}&= \sum_{i} Y^{(i)}_{2k-1}\otimes X^{(i)}_{2k}\\
    B_{2k,2k+1}&= \sum_{i} X^{(i)}_{2k}\otimes Z^{(i)}_{2k+1},
\end{split}
\ee
where the operators $Y^{(i)}_{2k-1}$ and $Z^{(i)}_{2k+1}$ are arbitrary. The condition \eqref{eq: conition even} implies 
\be
\begin{split}
\sum_{i_1,\dots, i_m} \bigotimes_{k=1}^m X^{(i_k)}_{2k} \bigotimes_{k} Y^{(i_k)}_{2k-1} =
\sum_{i_1,\dots, i_m} \bigotimes_{k=1}^m X^{(i_k)}_{2k} \bigotimes_{k} Z^{(i_k)}_{2k+1}.
\end{split}
\ee
Because the operators $X^{(i)}_{2k}$ are orthonormal it follows that
\be
\bigotimes_{k=1}^m Y^{(i_k)}_{2k-1} = \bigotimes_{k=1}^m Z^{(i_k)}_{2k+1}
\ee
for all values of the indices $i_1,\dots, i_m$. This implies that for fixed indices values the operators acting on the same system, e.g. $Y^{(i_{k+1})}_{2k+1}$ and $Z^{(i_{k})}_{2k+1}$, are equal up to a constant the operators are equal up to a multiplicative factor. Now, chose a value $i_k=j$ such that  $Z^{(j)}_{2k+1}\neq 0$, such a value exists because $B_{2k,2k+1}\neq 0$ by assumption. For any value $i_{k+1}$ we thus have  $Y_{2k+1}^{(i_{k+1})} = c_{2k+1}^{(i_k+1)} Z^{(j)}_{2k+1}$ for some scalar $c_{2k+1}^{(i_k+1)}$. In turn, since at least one $Y_{2k+1}^{(i_{k+1})}$ is nonzero, we obtain that $Z^{(i_{k})}_{2k+1} = \tilde c_{2k+1}^{(i_k)} Z^{(j)}_{2k+1}$. We conclude that for each system there is a nonzero operator $C_{2k+1}$ and scalar values $y_{ik}$ and $z_{ik}$ such that 
\be
Y^{(i)}_{2k-1}= y_{ik} C_{2k-1} \qquad Z^{(i)}_{2k+1}= z_{ik} C_{2k+1}
\ee
For the original operators in Eq.~\eqref{eq app: init A B} this gives
\be
\begin{split}
      A_{2k-1,2k}&= \sum_{i} y_{ik} C_{2k-1} \otimes X^{(i)}_{2k} = C_{2k-1}\otimes \left(\sum_{i} y_{ik} X^{(i)}_{2k}\right)\\
    B_{2k,2k+1}&= \sum_{i} X^{(i)}_{2k}\otimes z_{ik } C_{2k+1} =\left (\sum_{i} z_{ik }X^{(i)}_{2k}\right)\otimes C_{2k+1} .
    \end{split}
\ee
We have thus proven that all $A_{2k-1,2k}$ and $B_{2k,2k+1}$ are product.$\square$
\bigskip

\subsection{Proof of Lemma 1o}
\label{sec: proof 1o}

Consider $n=2m+1$ quantum systems, a set of $m+1$ bipartite nonzero operators $A_{1,2}, A_{3,4}, \dots, A_{2m-1,2m}, A_{2m+1, 1}$ and a set of $m$ operators $B_{2,3}, B_{4,5}, \dots, B_{2m,2m+1}$, which satisfy
\be
\left(\bigotimes_{k=1}^m A_{2k-1,2k} \right) A_{2m+1,1} = 
\bigotimes_{k=1}^m B_{2k,2k+1},
\ee
with the labels indicating on which pair of systems the operators act.

To show that the operators are products we again introduce  an orthonormal operator basis $\{X_{2k}^{(i)}\}$ for each even system. system. This allows us to express
\be\label{eq app: lemma odd A B}\begin{split} 
    A_{2k-1,2k}&= \sum_{i} Y^{(i)}_{2k-1}\otimes X^{(i)}_{2k}\quad k \leq m\\
    B_{2k,2k+1}&= \sum_{i} X^{(i)}_{2k}\otimes Z^{(i)}_{2k+1}.
\end{split}
\ee
We then rewrite the condition of the lemma as 
\be
\left(\sum_{i_1,\dots, i_m} \bigotimes_{k=1}^m X^{(i_k)}_{2k} \bigotimes_{k} Y^{(i_k)}_{2k-1}\right) A_{2m+1,1} 
=
\sum_{i_1,\dots, i_m} \bigotimes_{k=1}^m X^{(i_k)}_{2k} \bigotimes_{k} Z^{(i_k)}_{2k+1}.
\ee
By linear independence of all $X_{2k}^{(i)}$ this implies
\be\label{eq: odd cond 2}
\left(\bigotimes_{k=1}^m Y^{(i_k)}_{2k-1}\right)A_{2m+1,1} = \mathds{1}_1\bigotimes_{k=1}^m Z^{(i_k)}_{2k+1}.
\ee

Next, let us chose the values of the coefficients $i_1,\dots i_m$ such that the $Z^{(i_k)}_{2k+1}$ are all nonzero (we known that $B_{2k,2k+1}$ are nonzero hence such values exist), and relabel these values to $i_1=\dots=i_m=1$. We find
\be\label{eq app: whatever}
(Y_1^{(1)}\otimes Y_{3}^{(1)}\otimes \dots Y_{2m-1}^{(1)}\otimes \mathds{1}_{2m+1})A_{2m+1,1}
= \mathds{1}_1 \otimes Z_3^{(1)}\otimes \dots Z_{2m+1}^{(1)}
\ee
Now multiply both sides of the equation by $Z^{(1)\dag}_{3}\otimes \dots Z^{(1)\dag}_{2m+1}$ from te right and trace out all the systems but the first one. We find
\be
Y_1^{(1)} (\tr A_{2m+1,1} Z^{(1)\dag}_{2m+1}) \prod_{k=1}^{m-1} \tr Y^{(1)}_{2k+1}Z^{(1)\dag}_{2k+1} 
= \mathds{1}_1 \prod_{k=1}^{m} \tr Z^{(1)}_{2k+1}Z^{(1)\dag}_{2k+1} 
\ee
where the right hand side is nonzero since we have chosen the operators $Z^{(1)}_{2k+1}$ that are all nonzero. We can thus conclude that 
\be
Y_1^{(1)} (\tr A_{2m+1,1} Z^{(1)\dag}_{2m+1}) = \xi\,  \mathds{1}_1
\ee
with a nonzero constant $\xi \equiv \prod_{k=1}^{m} \tr Z^{(1)}_{2k+1}Z^{(1)\dag}_{2k+1}$. The last equation guarantees that the operator $Y_1^{(1)}$ is inevitable. Multiplying the Equation~\eqref{eq app: whatever} by  $(Y_1^{(1)})^{-1}$ we find 
\be
Y_{3}^{(1)}\otimes \dots Y_{2m-1}^{(1)}\otimes A_{2m+1,1}
=  Z_3^{(1)}\otimes \dots Z_{2m-1}^{(1)} \otimes Z_{2m+1}^{(1)} \otimes(Y_1^{(1)})^{-1},
\ee
and in particular
\be
A_{2m+1,1} = a\,  Z_{2m+1}^{(1)} \otimes (Y_1^{(1)})^{-1},
\ee
where we defined a nonzero constant $a$. This shows that $A_{2m+1,1}$ is product. 

Finally, the last identity allows us to rewrite the Eq.~\eqref{eq: odd cond 2} as
\be
a\, Y_{1}^{(i_1)} (Y_1^{(1)})^{-1} \bigotimes_{k=2}^m Y_{2k-1}^{(i_k)}\otimes Z^{(1)}_{2m+1}
=\mathds{1}_1\bigotimes_{k=1}^m Z^{(i_k)}_{2k+1},
\ee
again this guarantees that all the operators acting on the same system are equal up to a multiplicative factor. With identical arguments to those in the even lemme we arrive to 
\be
Y^{(i)}_{2k+1}= y_{ik} C_{2k+1} \qquad Z^{(i)}_{2k-1}= z_{ik} C_{2k-1}.
\ee
Plugging these relations in the Eq.~{eq app: lemma odd A B} we conclude that 
\be
\begin{split}
      A_{2k-1,2k}&= \sum_{i} y_{ik} C_{2k-1} \otimes X^{(i)}_{2k} = C_{2k-1}\otimes \left(\sum_{i} y_{ik} X^{(i)}_{2k}\right)\quad k<m\\
    B_{2k,2k+1}&= \sum_{i} X^{(i)}_{2k}\otimes Z_{ik } C_{2k+1} =\left (\sum_{i} z_{ik }X^{(i)}_{2k}\right)\otimes C_{2k+1}.
    \end{split}
\ee
Since we already know that $A_{2m+1,1}$ is product this concludes the proof. $\square$

\section{Appendix C. Proof of result 2}
\label{app: theorem 2}

\textbf{Result 2} \textit{Consider a quantum strategy on the $n$-partite ring network with the global state $\ket{\Psi}=\ket{\psi_1}_{R_1L_{2}}\dots \ket{\psi_n}_{R_n L_1}$ and the measurements given by POVMs $\{E_{X_k}^{a_k}\}_{a_k}$ acting on systems $L_k R_k$. If the strategy leads to a TC distribution $P(a_1,\dots a_n)$, arising steaming from a TC strategy with the $N_1, \dots N_n$ tokens distributed by each source accordingly to the probability distrubutions $p_1(t_1),\dots p_n(t_n)$, then each quantum system} $R_k=\textbf{R}_k J_{k}^R$\textit{ and }$L_k=\textbf{L}_k J_{k}^L$ \textit{can be decomposed in subsystems such that the quantum strategy takes the form}
\be\label{eq: TC quantum thrm}\begin{split}
\ket{\psi_k}_{R_k L_{k+1}}&=\sum_{t=0}^{N_k} \sqrt{p_k(t)} \ket{t,N_k-t}_{\textbf{R}_k\textbf{L}_{k+1}}\ket{j^{(t)}_k}_{J_{k}^R J_{k+1}^L}\\
E_{X_k}^{a_k}=\Pi_{X_k}^{a_k} &= \left(\sum_{t+t'=a_k} \ketbra{t}_{\textbf{L}_k} \otimes \ketbra{t'}_{\textbf{R}_{k}}\right)\otimes \mathds{1}_{J_k^L J_k^R} 
\end{split}
\ee

To each system $R_k(L_k)$ let us associate an Hilbert space $\cH_{R_k(L_k)}$ on which the state $\ket{\psi_k}(\ket{\psi_{k-1}})$ is supported, it is assumed to have an arbitrary but finite dimension. Next, by introduction an auxiliary system $M_k$ in the state $\ket{0}\in \cH_{M_k}$ we dilate each POVMs $\{E_{X_k}^{a_k}\}$ to a PVM $\{ \bar \Pi_{X_k}^{a_k}\}$, with the projectors $\bar \Pi_{X_k}^{a_k}$ acting on the systems $R_kL_kM_k$ and
\be\label{app: POVM}
E_{X_k}^{a_k} = \tr_{M_k} \bar \Pi_{X_k}^{a_k} (\mathds{1}_{L_kR_K}\otimes \ketbra{0}_{M_k}) = \bra{0}_{M_k}  \bar \Pi_{X_k}^{a_k} \ket{0}_{M_k}
\ee
For short we collect all the states of the axilliary systems into $\ket{\bm 0}=\ket{0}_{M_1}\dots \ket{0}_{M_n}$.

Following the main text, this allows us to define a unitary operator for each party $k$
\be\label{eq: app def Uk}
\bar U_k = \sum_{x=0}^N e^{\ii \varphi_x}\, \bar \Pi_{X_k}^x \qquad \text{with} \qquad e^{\ii \varphi_x} = \exp(\ii (x+\frac{1}{n})\frac{2\pi}{N+1})
\ee
Note that in general not all outcomes $x=a_k \in [0,\dots,N]$ are possible, hence some of the projectors 
$\{\bar \Pi_{X_k}^{x}\}_{x}$ can be zero. The TC correlations satisfy 
\be
a_1+\dots + a_n = N.
\ee
For our quantum model   it implies 
\be
\bar \Pi^{a_1}_{X_1} \dots \bar \Pi^{a_n}_{X_n} \ket{\Psi}\ket{\bm 0}=0 \quad \text{if} \quad a_1+\dots + a_n \neq N
\ee
and  guarantees that 
\be\label{eq: app cons dilate}
\bigotimes_{k=1}^n \bar U_k \ket{\Psi}\ket{\bm 0}=\ket{\Psi}\ket{\bm 0}.
\ee
To see this expand the unitaries as $\bigotimes_{k=1}^n \bar U_k= \sum_{a_1\dots a_n}\exp(\ii  \frac{1+\sum_{i=1}^n a_i }{N+1}2\pi )\bar \Pi^{a_1}_{X_1} \dots \bar \Pi^{a_n}_{X_n} $ and note that
\be
\exp(\ii  \frac{1+\sum_{i=1}^n a_i }{N+1}2\pi )\bar \Pi^{a_1}_{X_1} \dots \bar \Pi^{a_n}_{X_n} \ket{\Psi}\ket{\bm 0}
\ee
is zero unless $a_1+\dots + a_n = N $, or $\exp(\ii \frac{1+\sum_{i=1}^n a_i }{N+1} 2\pi)=1$. Therefore $\ket{\Psi}\ket{\bm 0}$ is an eigenstate of $\bigotimes_{k=1}^n \bar U_k $ with eigenvalue 1.

Applying our result 0 to Eq.~\eqref{eq: app cons dilate} gurantees that the original measurements $\{E_{X_k}^{a_k} = \Pi_{X_k}^{a_k}\}_{a_k}$ are projective. We can thus rewrite the constraint~\eqref{eq: app cons dilate} in a simpler form 
\begin{align}    
\label{eq: app cons not dilate}
&\bigotimes_{k=1}^n  U_k \ket{\Psi}=\ket{\Psi} \qquad \text{with}\\
& \sum_{x=0}^N \exp(\ii (x+\frac{1}{n})\frac{2\pi}{N+1})\, \Pi_{X_k}^x.
\end{align}
Result 1 then implies that all the unitaries $U_X$ are product
\be
U_X = \sum_{x=0}^N \exp(\ii (x+\frac{1}{n})\frac{2\pi}{N+1})\, \Pi_{X_k}^x = V_{L_k}^{(k)}\otimes W_{R_k}^{(k)}.
\ee

Let us now focus on the eigenvalues of the unitaries $V_{L_k}^{(k)}$ and $W_{R_k}^{(k)}$, which we know have to fulfill the identity
\be\label{eq: eigenvalues app}
\exp\big(\ii (v_j+w_\ell)\big) = \exp(\ii (a_k+\frac{1}{n})\frac{2\pi}{N+1})
\ee
for some $a_k$. That is, for each pair of eigenvalues with $v_j$ and $w_\ell$ there is an $a_k$ fulfilling Eq.~\eqref{eq: eigenvalues app}, and conversely for each possible output value $a_k$ there is a a pair $v_i$ and $w_j$ satisfying Eq.~\eqref{eq: eigenvalues app}.
By adjusting the relative phase of the unitaries in the decomposition~\eqref{eq: eigenvalues app}, $(V_{L_k}^{(k)},W_{R_k}^{(k)}) \to (e^{\ii \omega}V_{L_k}^{(k)},e^{-\ii \omega} W_{R_k}^{(k)})$ we are free to shift the eigenvalues $v_j\to v_j +\omega$ and $w_j \to -\omega$. Let us pick a decomposition such that $V_{L_k}^{(k)}$ admits the eigenvalue $e^{\ii v_0}=1$, and ask what are the possible values of $e^{\ii w_j}$.
Form $e^{\ii (v_0+w_\ell)}= e^{\ii (a_k+\frac{1}{n}) \frac{2\pi}{n(N+1)}}$
we conclude that the possible values are $e^{\ii w_\ell} = e^{\ii \frac{2\pi}{n(N+1)}}e^{\ii a_k \frac{2\pi}{N+1}}$ for some possible value of $a_k$. In turn, the same argument with any of the $w_\ell$ guarantees the possible values of $e^{\ii v_j}$ are of the form $e^{\ii a_k \frac{2\pi}{N+1}}$. Both can take at most $N+1$ values. We thus obtain a decomposition $U_{k} = V_{L_k}^{(k)}\otimes W_{R_k}^{(k)}$ with
\be\begin{split}
V_{L_k}^{(k)} &= \sum_{j=0}^N e^{\ii j \frac{2\pi}{N+1}} \Pi_{L_k}^j \\
W_{R_k}^{(k)} &= e^{\ii\frac{2\pi}{n(N+1)}} \sum_{\ell=0}^N e^{\ii \ell \frac{2\pi}{N+1}} \Pi_{R_k}^\ell.
\end{split}\ee
Here again, some of the projectors can be zero because not all of the values $j$ and $\ell$ are generally possible.

In any case this decomposition allows us to split the Hilbert space associated with each system as e.g.
\be
\cH_{L_k} = \bigoplus_{j=0}^N \cH_{L_k}^{(j)},
\ee
where each  $\cH_{L_k}^{(j)}$ is the subspace on which $\Pi_{L_k}^j$ projects. It is more insightful to write this decomposition as a tensor product instead of a direct sum. While the subspaces $\cH_{L_k}^{(j)}$ might have different dimensions, it is always possible to complete them with virtual levels (that do not support $\ket{\Psi}$) to make their dimensions match. Then we can write 
\be\label{eq: app decomposition tensor}
\cH_{L_k} = \mathds{C}^{N+1}_{\textbf{L}_k} \otimes \cH_{J_{k}^L},
\ee
where $\mathds{C}^{N+1}_{\textbf{L}_k}$ describes the quit ($d=N+1$) that carries the value $j$, and $\cH_{J_{k}^L}$ collects all the other degrees of freedom necessary to describe $L_k$ but that do not influence the measurement outcome. The same logic can be applied to the system $R_k$. By construction we obtain
\be\begin{split}
V_{L_k}^{(k)} &= \left(\sum_{j=0}^{N} e^{\ii j \frac{2\pi}{N+1}} \ketbra{j}_{\textbf{L}_k}\right) \otimes \mathds{1}_{J_{k}^L} \\
W_{R_k}^{(k)}&=  e^{\ii\frac{2\pi}{n(N+1)}} \left(\sum_{\ell=0}^{N} e^{\ii \ell \frac{2\pi}{N+1}} \ketbra{\ell}_{\textbf{R}_k}\right) \otimes \mathds{1}_{J_{k}^R}.    
\end{split}
\ee
Combining them, we obtain 
\be
U_k = \left(  \sum_{j,\ell=0}^N e^{\ii (j+\ell +\frac{1}{n})\frac{2\pi }{N+1}} \ketbra{j,\ell}_{\textbf{L}_k \textbf{R}_k}\right)\otimes \mathds{1}_{J_{k}^{X}}.
\ee
Finally, comparing with the definition~\eqref{eq: app def Uk} we obtain a decomposition of the measurement operators
\be\label{eq: app Pi first}
\Pi_{X_k}^{a_k} =  \left(  \sum_{\substack{j+\ell \equiv a_k \\(\text{mod} N+1)}} \ketbra{j,\ell}_{\textbf{L}_k \textbf{R}_k}\right)\otimes \mathds{1}_{J_{k}^X}.
\ee

The decomposition~\eqref{eq: app decomposition tensor} can also be used to express the states prepared by the source as 
\be\label{eq: app state first}
\ket{\psi_k} = \sum_{i,j=0}^N \Psi^{(k)}_{ij} \ket{ij}_{\textbf{R}_k \textbf{L}_{k+1}} \ket{j_k^{(ij)}}_{J_{k}^R J_{k+1}^L}.
\ee
Here, the states of the junk systems  $\ket{j_k^{(ij)}}_{J_{k}^R J_{k+1}^L}$ have no influence on the measurement outcomes and are completely arbitrary. Without loss of generality the amplitudes $\Psi_{ij}^{(k)}$ can be taken real positive, by absorbing any complex phase inside $\ket{j_k^{(ij)}}_{J_{k}^R J_{k+1}^L}$.

So far we have only used the equality $\sum_k a_k =N$ to derive rather strict restrictions~(\ref{eq: app Pi first},\ref{eq: app state first}) on the form of any the quantum model leading to a TC distribution.
Let us now use the classical rigidity property TC distributions $P(a_1,\dots,a_n)$ to show that amplitudes $\Psi^{(k)}_{ij}$ are essentially unique. Recall that a TC distribution (strategy) are characterized by a fixed number of tokens $N_1,\dots ,N_n$ and the probability distribution $p_k(i)$ that the source $S_k$ send $i$ tokens to $R_k$ and $N_k-i$ tokens to $L_{k+1}$ for $i\in\{0,\dots,N_k\}$. 

First, remark  that the equations~(\ref{eq: app Pi first},\ref{eq: app state first}) also define a unique classical strategy. Each source $S_k$ samples a pair of integers $(i,j) \in\{0,\dots,N\}^{\times 2}$  from the probability distribution 
\be
\text{P}^{(k)}(i,j)=|\Psi_{ij}^{(k)}|^2,
\ee
the value $i$ and $j$ define the states of the classical systems $R_k$ and $L_{k+1}$ sent to the neighbouring parties. The party $k$ reads the values $j$ from  $L_k$ and $\ell$ for $R_k$, and outputs
\be
a_k(j,\ell) = (j + \ell) \, \text{mod}\,  (N+1).
\ee
The rigidity of classical TC strategies~\cite{Renou_rigid} guarantees that for all systems $R_k$ and $L_k$ there exist "token functions"
\be\begin{split}
T_k^R: \{0,\dots,N\} &\to \{0,\dots,N_k\} \\
T_{k+1}^L: \{0,\dots,N\} &\to \{0,\dots,N_k\}    
\end{split}
\ee
such that 
\begin{itemize}
    \item[(i)] $T_k^{R}(i)+T_{k+1}^L(j)=N_k$ if $P^{(k)}(i,j)\neq 0$. \\
    \item[(ii)] $a_k(j,\ell)=T_{k-1}^{L}(j) +T_{k}^{R}(\ell)$ for all possible values $j$ and $\ell$ (nonzero probability).
    \item[(iii)] The tokens are distributed in the same way that in the TC strategy. That is
    \be
  \sum_{i,j} \text{P}^{(k)}(i,j) \delta_{T_k^R(i), i'} \delta_{T_{k+1}^L(j), N_k-i'} = p_k(i').
    \ee
\end{itemize}  

For each system the token function define disjoint subsets $\mathds{T}^t_{R_k(L_k)}\subset \{0,\dots,N\}$, such that $T_{k}^R(i) = t$ if $i\in \mathds{T}^t_{R_k}$. For a quantum model this defines a block diagonal structure for our qudits $\mathds{C}^{d}_{\textbf{R}_k}$ and  $\mathds{C}^{d}_{\textbf{L}_{k+1}}$ in $
N_{k}+1$ blocks. As before we can embed the qudits $\mathds{C}^{d}_{\textbf{R}_k}$, $\mathds{C}^{d}_{\textbf{L}_{k+1}}$ into tensor product spaces $\mathds{C}^{(N_k+1)}_{\textbf{T}^R_k}\otimes \mathcal{M}_{R_k}$ and  $\mathds{C}^{(N_k+1)}_{\textbf{T}^L_{k+1}}\otimes \mathcal{M}_{L_{k+1}}$ in order to write
\be\begin{split}
\ket{i}_{\textbf{R}_k} &= 
\ket{t_i=T_k^{R}(i)}_{\textbf{T}^R_{k}} \ket{\alpha_i}_{M_{k}^R} \\
\ket{j}_{\textbf{L}_{k+1}} &= \ket{t_j=T_{k+1}^{L}(j)}_{\textbf{T}^L_{k+1}} \ket{\alpha_j}_{M_{k+1}^L} 
\end{split}
\ee

By the property (ii) we known that the response functions $a_k(j,\ell) = t_j +t_\ell$ only depend on the the values $t_j$ and $t_\ell$ but not on the multiplicities. Hence, we can rewire Eq.~\eqref{eq: app Pi first} as
\be
\Pi_{X_k}^{a_k} =  \left(  \sum_{t+t'=a_k} \ketbra{t,t'}_{\textbf{T}_k^L\textbf{T}_k^R}\right)\otimes \mathds{1}_{M_k^L M_k^R}\otimes \mathds{1}_{J_{k}^X}.
\ee
In turn the state of Eq.~\eqref{eq: app state first} can be rewritten as 
\be\label{eq: app state whatever}\begin{split}
\ket{\psi_k} &= \sum_{i,j=0}^N \Psi^{(k)}_{ij} \ket{ij}_{\textbf{R}_k \textbf{L}_{k+1}} \ket{j_k^{(ij)}}_{J_{k}^R J_{k+1}^L}\\
& = \sum_{i,j=0}^N \Psi^{(k)}_{ij} \ket{t_i,t_j}_{\textbf{T}_k^R \textbf{T}_{k+1}^L}\ket{\alpha_i,\alpha_j}_{M_k^R M_{k+1}^{L}} \ket{j_k^{(ij)}}_{J_{k}^R J_{k+1}^L}
\\
&=\sum_{t,t'=0}^{N_k} \ket{t,t'}_{\textbf{T}_k^R\textbf{T}_{k+1}^L}
 \underbrace{\sum_{i\in\mathds{T}_{R_k}^t}\sum_{ j\in\mathds{T}_{L_{k+1}}^{t'}} \Psi^{(k)}_{ij} \ket{\alpha_i,\alpha_j}_{M_k^R M_{k+1}^L} \ket{j_k^{(ij)}}_{J_{k}^R J_{k+1}^L}}_{\equiv \Gamma_{t,t'} \ket{j_k^{(t,t')}}_{M_k^R M_{k+1}^LJ_{k}^R J_{k+1}^L}}.
\end{split}
\ee
Compute the normalization of the state defined in the last line
\be\begin{split}
   | \Gamma_{t,t'} |^2 =
   &  \|\sum_{i\in\mathds{T}_{R_k}^t}\sum_{ j\in\mathds{T}_{L_{k+1}}^{t'}} \Psi^{(k)}_{ij} \ket{\alpha_i,\alpha_j}_{M_k^R M_{k+1}^L} \ket{j_k^{(ij)}}_{J_{k}^R J_{k+1}^L} \|^2 \\
   & = \sum_{i\in\mathds{T}_{R_k}^t}\sum_{ j\in\mathds{T}_{L_{k+1}}^{t'}} |\Psi^{(k)}_{ij}|^2 \\
   &= \sum_{i\in\mathds{T}_{R_k}^t}\sum_{ j\in\mathds{T}_{L_{k+1}}^{t'}} \text{P}^{(k)}(i,j)\\
   & = p_k(t)
\end{split}
\ee
by (iii). Furthermore, by (i) we know that in Eq.~\eqref{eq: app state whatever} $t_i+t_j =N_k$ whenever $\Psi^{(k)}_{ij}$ is nonzero. Therefore,
\be
\ket{\psi_k}= \sum_{t=0}^{N_k} \sqrt{p_k(t)} \ket{t,N_k-t}_{\textbf{T}_k^R\textbf{T}_{k+1}^L} \ket{j_k^{(t,t')}}_{M_k^R M_{k+1}^LJ_{k}^R J_{k+1}^L}.
\ee
It remains to absorb the systems $M_k^R$ and $M_{k+1}^{L}$ inside the junk to obtain the desired decomposition
\be\begin{split}
    \ket{\psi_k}&= \sum_{t=0}^{N_k} \sqrt{p_k(t)} \ket{t,N_k-t}_{\textbf{T}_k^R\textbf{T}_{k+1}^L} \ket{j_k^{(t)}}_{J_{k}^R J_{k+1}^L} \\
    \Pi_{X_k}^{a_k} &=  \left(  \sum_{t+t'=a_k} \ketbra{t,t'}_{\textbf{T}_k^L\textbf{T}_k^R}\right)\otimes\mathds{1}_{J_{k}^R J_{k+1}^L}. \square
\end{split}
\ee

\section{Appendix D. Derivation of the bound on the coherence $r$ of quantum models simulating the RGB4 distribution.}

\subsection{Dilation of the measurements $\{\Pi_X^{0},E_X^{1_0},E_X^{1_1},\Pi_X^{2}\}$} 
We start by commenting on the dilation needed to write $\Pi_X^{1_i}$ as projectors. Note that result 2 only guarantees that the coarse-grained measurements are projective
\be
\Pi_X^0 = \ketbra{00}_{\textbf{X}_\xi \textbf{X}_{\xi'}}\otimes \mathds{1}_{J_X},\quad 
\Pi_X^2=\ketbra{11}_{\textbf{X}_\xi \textbf{X}_{\xi'}}\otimes \mathds{1}_{J_X}, \quad
 \Pi_X^1=(\ketbra{01}+\ketbra{10})_{\textbf{X}_\xi \textbf{X}_{\xi'}}\otimes \mathds{1}_{J_X}.
\ee
This is not necessarily the case of the operators $E_{X}^{1_0}$ and $E_{X}^{1_1}$, that only need to satisfy
\be
E_{X}^{1_0}+E_{X}^{1_1} = \Pi_X^{1}. 
\ee
Nevertheless, we these measurements can be dilated to projectors For the sake of clarity we present it explicitly here. First, diagonalize the operators 
\be
E_X^{1_0} = \sum_l p_k \ketbra{\phi_k} \qquad E_X^{1_1} = \sum_l (1-p_k) \ketbra{\phi_k}
\ee
where the states $\{\ket{\phi_k}\}_k$ define a basis of the Hilbert space $\cH_X^{(1)}=span\left(\ket{01}_{\textbf{X}_\xi \textbf{X}_{\xi'}},\ket{10}_{\textbf{X}_\xi \textbf{X}_{\xi'}} \right) \otimes \cH_{J_k^L}\otimes \cH_{J_k^R}$. Next, introduce an auxiliary qubit  $M_X$ initially in the state $\ket{0}$, and define a unitary 
\be
V = \sum_{k} (\sqrt{p_k}\ket{\phi_k,0}+ \sqrt{1-p_k} \ket{\phi_k,1})\bra{\phi_k,0} + (\sqrt{1-p_k}\ket{\phi_k,0}- \sqrt{p_k} \ket{\phi_k,1})\bra{\phi_k,1}
\ee
acting on $\cH_X^{(1)}\otimes\mathds{C}^2_{M_X}$, and the two projectors
\be\begin{split}
\bar \Pi_X^{1_0} &= \bra{0}V^\dag (\Pi_X^1\otimes \ketbra{0}_{M_X}) V \ket{0}\\
\bar \Pi_X^{1_1} &= \bra{0} V^\dag (\Pi_X^1\otimes \ketbra{1}_{M_X}) V \ket{0}.
\end{split}
\ee
One verifies that this is indeed a dilation of the original operators $E_X^{1_i}= \bra{0} \bar \Pi_X^{1_i}  \ket{0}$, as
\be\begin{split}
(\Pi_X^1\otimes \ketbra{0}_{M_X}) V\ket{0} &= \sum_{k} \sqrt{p_k}\ketbra{\phi_k,0}{\phi_k}
\\
(\Pi_X^1\otimes \ketbra{1}_{M_X}) V\ket{0} &= \sum_{k} \sqrt{p_k}\ketbra{\phi_k,1}{\phi_k}.
\end{split}
\ee
Furthermore, 
\be
\bar{\Pi}_X^1 = \bar \Pi_X^{1_0}+\bar \Pi_X^{1_1} = V^\dag (\Pi_X^{1}\otimes\mathds{1}_{M_X}) V= \Pi_X^{1}\otimes\mathds{1}_{M_X} 
\ee
Finally, we get the complete PVM $\{\bar \Pi_X^{0},\bar \Pi_X^{1_0},\bar \Pi_X^{1_1},\bar \Pi_X^{2}\}$ by defining
\be
\bar \Pi_X^{0} = \Pi_X^{0}\otimes \mathds{1}_{M_X} \qquad \bar \Pi_X^{2} = \Pi_X^{2}\otimes \mathds{1}_{M_X}.
\ee
So that
\be\begin{split}
\bar \Pi_X^0 &= \ketbra{00}_{\textbf{X}_\xi \textbf{X}_{\xi'}}\otimes \mathds{1}_{J_X}\otimes \mathds{1}_{M_X},\quad 
\bar \Pi_X^2=\ketbra{11}_{\textbf{X}_\xi \textbf{X}_{\xi'}}\otimes \mathds{1}_{J_X}\otimes \mathds{1}_{M_X}\\
 \bar \Pi_X^1&=\bar \Pi_X^{1_0}+\bar \Pi_X^{1_1} = (\ketbra{01}+\ketbra{10})_{\textbf{X}_\xi \textbf{X}_{\xi'}}\otimes \mathds{1}_{J_X}\otimes \mathds{1}_{M_X}.
\end{split}
\ee

Let us now simply absorb each qubit $M_X$ into one of the junk system, say the one received by the party from the left $\bar J_k^L = J_k^L M_X$, and $\ket{j_\xi^{c(a)}}_{\bar{J}\xi} = \ket{j_\xi^{c(a)}}_{J_\xi}\ket{0}_{M_X}$. We obtain the states
\be
 \ket{\psi_\xi}= \frac{1}{\sqrt{2}}\big( \ket{01}_{\textbf{X}_\xi \textbf{Y}_\xi}\!\ket{j^{c}_\xi}_{\bar J_\xi}\!\!+
\ket{10}_{\textbf{X}_\xi \textbf{Y}_\xi}\ket{j^{a}_\xi}_{\bar J_\xi},
 \big)
\ee
and $\ket{\Psi}= \ket{\psi_\alpha}\ket{\psi_\beta}\ket{\psi_\gamma}$
such that $P_Q(a,b,c)= \|\bar \Pi_A^a \bar \Pi_B^b \bar \Pi_C^c \ket{\Psi}\|^2$
with projective measurements. At this point we forget the bars over the projectors and the junk systems, and simply write 
\be
P_Q(a,b,c)= \| \Pi_A^a  \Pi_B^b  \Pi_C^c \ket{\Psi}\|^2.
\ee

\subsection{Derivation of the bound for $r$}

Let us now compute the probabilities of the equation \eqref{eq: cons projectors} in the main text, starting with $P_Q(1_i,1_j,1_k) $. Since $\Pi^{1_i}_A \Pi^{1_j}_B\Pi^{1_k}_C=(\Pi^{1_i}_A \Pi^{1_j}_B\Pi^{1_k}_C)^2$ are projectors we express 
\be\begin{split}
   P_Q(1_i,1_j,1_k) &=\| \Pi^{1_i}_A \Pi^{1_j}_B\Pi^{1_k}_C \ket{\Psi}\|^2 \\
   &=\frac{1}{8} \| \Pi^{1_i}_A \Pi^{1_j}_B\Pi^{1_k}_C \big( \ket{01}_{\textbf{B}_\alpha \textbf{C}_\alpha}\!\ket{j^{c}_\alpha}_{J_\alpha}\!\!+
\ket{10}_{\textbf{B}_\alpha \textbf{C}_\alpha}\ket{j^{a}_\alpha}_{J_\alpha}
 \big) \otimes \\ & \qquad
 \big( \ket{01}_{\textbf{C}_\beta \textbf{A}_\beta}\!\ket{j^{c}_\beta}_{J_\beta}\!\!+
\ket{10}_{\textbf{C}_\beta \textbf{A}_\beta}\ket{j^{a}_\beta}_{J_\beta}
 \big)\otimes
 \big( \ket{01}_{\textbf{A}_\gamma \textbf{B}_\gamma}\!\ket{j^{c}_\gamma}_{J_\gamma}\!\!+
\ket{10}_{\textbf{A}_\gamma \textbf{B}_\gamma}\ket{j^{a}_\gamma}_{J_\gamma}
 \big)\|^2.
\end{split}
\ee
Each $\Pi_{X}^{1_x}$ is only supported on the subspace where the the party receives a single token $\text{span}\{\ket{01}_{X_\xi X_{\xi'}},\ket{10}_{X_\xi X_{\xi'}}\}$, therefore in the above expression only two terms are nonzero.
\be\begin{split}
   &P_Q(1_i,1_j,1_k) = \\
   &=\frac{1}{8} \| \Pi^{1_i}_A \Pi^{1_j}_B\Pi^{1_k}_C \big( \ket{01}_{\textbf{B}_\alpha \textbf{C}_\alpha}\!\ket{j^{c}_\alpha}_{J_\alpha}\!\!\ket{01}_{\textbf{C}_\beta \textbf{A}_\beta}\!\ket{j^{c}_\beta}_{J_\beta}\!\! \ket{01}_{\textbf{A}_\gamma \textbf{B}_\gamma}\!\ket{j^{c}_\gamma}_{J_\gamma}\!\!
   +
\ket{10}_{\textbf{B}_\alpha \textbf{C}_\alpha}\ket{j^{a}_\alpha}_{J_\alpha}\!\! \ket{10}_{\textbf{C}_\beta \textbf{A}_\beta}\ket{j^{a}_\beta}_{J_\beta}\!\!\ket{10}_{\textbf{A}_\gamma \textbf{B}_\gamma}\ket{j^{a}_\gamma}_{J_\gamma}
 \big)\|^2\\
 & =\frac{1}{8} \| \Pi^{1_i}_A \Pi^{1_j}_B\Pi^{1_k}_C \big(\ket{\Psi^c}+\ket{\Psi^a} \big)\|^2
\end{split}
\ee
where we defined the two global states $\ket{\Psi^c}\equiv\ket{01,01,01}_{\textbf{B}_\alpha \textbf{C}_\alpha \textbf{C}_\beta \textbf{A}_\beta  \textbf{A}_\gamma  \textbf{B}_\gamma}\ket{j_\alpha^c,j_\beta^c,j_\gamma^c}_{J_\alpha J_\beta J_\gamma}$ and $\ket{\Psi^a} \equiv \ket{10,10,10}_{\textbf{B}_\alpha \textbf{C}_\alpha \textbf{C}_\beta \textbf{A}_\beta  \textbf{A}_\gamma  \textbf{B}_\gamma}\ket{j_\alpha^a,j_\beta^a,j_\gamma^a}_{J_\alpha J_\beta J_\gamma}$. A priori, these are arbitrary quantum states
\be\begin{split}
\ket{\Psi^c} &= \underbrace{(\ket{j_\alpha^c}_{J_\alpha}\ket{01}_{\textbf{B}_\alpha \textbf{C}_\alpha} )}_{=\ket{\tilde{\psi}^c_\alpha}} \underbrace{(\ket{j_\beta^c}_{J_\beta}\ket{01}_{\textbf{C}_\beta \textbf{A}_\beta} )}_{=\ket{\tilde{\psi}^c_\beta}} \underbrace{(\ket{j_\gamma^c}_{J_\gamma}\ket{01}_{\textbf{A}_\gamma \textbf{B}_\gamma} )}_{=\ket{\tilde{\psi}^c_\gamma}}\\
\ket{\Psi^a} &= \underbrace{(\ket{j_\alpha^a}_{J_\alpha}\ket{10}_{\textbf{B}_\alpha \textbf{C}_\alpha} )}_{=\ket{\tilde{\psi}^a_\alpha}} \underbrace{(\ket{j_\beta^a}_{J_\beta}\ket{10}_{\textbf{C}_\beta \textbf{A}_\beta} )}_{=\ket{\tilde{\psi}^a_\beta}} \underbrace{(\ket{j_\gamma^a}_{J_\gamma}\ket{10}_{\textbf{A}_\gamma \textbf{B}_\gamma} )}_{=\ket{\tilde{\psi}^a_\gamma}}
\end{split}
\ee
on a triangle network, with the important property that for each system  $X_\xi$ the state $\ket{\Psi^a}$ and $\ket{\Psi^c}$ live on orthogonal subspaces (labeled by the state of the qubit $\textbf{X}_\xi$). With the help of the coherence $r = (-1)^{i+j+k} \,  2\, \text{Re}\bra{\Psi^c} \Pi^{1_i}_A \Pi^{1_j}_B\Pi^{1_k}_C \ket{\Psi^a}$ introduced in the main text, we obtain
\be
P_Q(1_i,1_j,1_k) =  \frac{1}{8} \left(\| \Pi^{1_i}_A \Pi^{1_j}_B\Pi^{1_k}_C \ket{\Psi^c}\|^2 +\| \Pi^{1_i}_A \Pi^{1_j}_B\Pi^{1_k}_C \ket{\Psi^a} \|^2 + (-1)^{i+j+k} r\right).
\ee

Next, let us compute $ P_Q(1_i,0,2)$. We obtain have
\be
\begin{split}
    P_Q(1_i,0,2) &= \| \Pi^{1_i}_A \Pi^{0}_B\Pi^{2}_C \ket{\Psi}\|^2 \\
    &=\| \Pi_A^{1_i} (\ketbra{00}_{\textbf{B}_\alpha \textbf{B}_\gamma}\otimes \mathds{1}_{J_B})(\ketbra{11}_{\textbf{C}_\alpha \textbf{C}_\beta}\otimes \mathds{1}_{J_C})\ket{\Psi}\|^2 \\
    &=\frac{1}{8}\| \Pi^{1_i}_A \ket{01,10,10}_{\textbf{B}_\alpha \textbf{C}_\alpha \textbf{C}_\beta \textbf{A}_\beta  \textbf{A}_\gamma  \textbf{B}_\gamma}\ket{j_\alpha^c,j_\beta^a,j_\gamma^a}_{J_\alpha J_\beta J_\gamma}\|^2 \\
    &= \frac{1}{8}\| \Pi^{1_i}_A \ket{01}_{\textbf{A}_\beta  \textbf{A}_\gamma}\ket{j_\beta^a,j_\gamma^a}_{J_\beta J_\gamma}\|^2\\
    & =\frac{1}{8}\| \Pi^{1_i}_A \ket{\Psi^a}\|^2
\end{split}
\ee
In the same way we obtain $P_Q(1_i,2,0) =\frac{1}{8}\|\Pi^{1_i}_A \ket{\Psi^c}\|^ 2$, and any cyclic permutation of these two equations. We obtain
\be\label{eq app: probas vu}\begin{split}
\|\Pi^{1_i}_A \Pi^{1_j}_B\Pi^{1_k}_C\ket{\Psi^c}\|^2\! +
\|\Pi^{1_i}_A \Pi^{1_j}_B\Pi^{1_k}_C\ket{\Psi^a}\|^2 +(-1)^{i+j+k} r &= (u_i u_j u_k + v_i v_j v_k)^2
\\   
\| \Pi^{1_i}_X \ket{\Psi^a}\|^2 & =u_i^2 \\
\|\Pi^{1_i}_X \ket{\Psi^c}\|^ 2 & = v_i^2.
\end{split}
\ee

Remark that the state $\ket{\Psi^c}$ and $\ket{\Psi^a}$ belong to the subspace where each party receive a single token, i.e. $(\Pi_X^{1_0}+\Pi_X^{1_1})\ket{\Psi^{c(a)}}=\ket{\Psi^{c(a)}}$. In this subspace, each pair of projectors $\{\Pi_X^{1_0},\Pi_X^{1_1}\}$ defines a PVM for the corresponding party. Accordingly, let us define the probability distributions
\be
q_c(i,j,k)=\|\Pi^{1_i}_A \Pi^{1_j}_B\Pi^{1_k}_C\ket{\Psi^c}\|^2
\qquad 
q_a(i,j,k)=\|\Pi^{1_i}_A \Pi^{1_j}_B\Pi^{1_k}_C\ket{\Psi^a}\|^2,
\ee
that both describe some quantum correlations on the triangle with binary outputs. These distributions have to satisfy $q_c(i,j,k)+q_a(i,j,k)+ (-1)^{i+j+k} =(u_i u_j u_k + v_i v_j v_k)^2$ and $q_c(i)=\sum_{jk} q_c(i,j,k)=v_i^2$ and $q_a(i)=\sum_{jk} q_a(i,j,k)=u_i^2$ by virtue of Eqs.~\eqref{eq app: probas vu}.

Following \cite{Renou_2019}, we also define the symmetrized distributions
\be\begin{split}
\tilde q_a(i,j,k) &= \frac{1}{6}(q_a(i,j,k)+q_a(j,k,i)+ q_a(k,i,j)+q_a(i,k,j)+q_a(k,j,i)+q_a(j,i,k))
    \\
\tilde q_c(i,j,k) & =  \frac{1}{6}(q_c(i,j,k)+q_c(j,k,i)+ q_c(k,i,j)+q_c(i,k,j)+q_c(k,j,i)+q_c(j,i,k))
\end{split}
\ee
The symmetrized distributions no longer describe quantum correlations on the triangle, as the implementation of the symmetrization procedure would require global shared randomness, nevertheless by convexity they are valid probability distributions and should satisfy the same constraints as the original distribution
\be\begin{split}
    \tilde q_c(i,j,k) +\tilde q_a(i,j,k)+(-1)^{i+j+k} r &=(u_i u_j u_k + v_i v_j v_k)^2\\
    \tilde q_c(i)&= \sum_{jk}\tilde q_c(i,j,k)=v_i^2 
    \\
    \tilde q_a(i)&= \sum_{jk}\tilde q_a(i,j,k)=u_i^2. 
\end{split}
\ee
Now let us also define
\be \begin{split}
\xi_{ijk} &= \frac{1}{2}(\tilde q_a(i,j,k) -\tilde q_c(i,j,k))\\
    \tilde q(i,j,k) &= \frac{1}{2} (\tilde q_a(i,j,k) +\tilde q_c(i,j,k))
\end{split}
\ee
with the short notation $\xi_x= \xi_{ijk}$ for $i+j+k=x$ (all of them are equal by symmetrization). Note that the average distribution $\tilde q(i,j,k)$ satisfies
\be
 \tilde q(i,j,k)  =\frac{(u_i u_j u_k + v_i v_j v_k)^2 - (-1)^{i+j+k} r}{2}.
\ee
or
\be
    q_a(i,j,k) = \tilde q(i,j,k) +\xi_{ijk}\qquad
    q_c(i,j,k) = \tilde q(i,j,k) -\xi_{ijk}.
\ee
Now we have the following equities
\be\begin{split}
\tilde q_a(i)&=\sum_{jk} \tilde q_a(i,j,k) =\sum_{jk} (\tilde q(i,j,k) +\xi_{i,j,k})\\
&= \frac{1}{2}\sum_{jk}((u_i u_j u_k + v_i v_j v_k)^2-r (-1)^{i+j+k}) +\sum_{jk}\xi_{ijk} \\
&= \sum_{jk}\frac{(u_i u_j u_k + v_i v_j v_k)^2}{2} +\sum_{jk}\xi_{ijk} \\
& = \frac{u_i^2+v_i^2}{2}+\sum_{jk}\xi_{ijk}\\
& = \frac{1}{2}+\sum_{jk}\xi_{ijk}.
\end{split}
\ee
Similarly, we have $q_c(i)= \frac{1}{2} -\sum_{jk}\xi_{ijk}.$\\

Using $\sum_{jk}\xi_{0jk} = \xi_0+2 \xi_1 +\xi_2$ and $\sum_{jk}\xi_{1jk} = \xi_1+2 \xi_2 +\xi_3$ we rewrite the above conditions together with the probability constraints as
\be\begin{split}
u^2 &= \frac{1}{2} + \xi_0+2 \xi_1 +\xi_2\\
1-u^2 &= \frac{1}{2} + \xi_1+2 \xi_2 +\xi_3.
\end{split}
\ee
That we use to write down
\be
\begin{split}
    \xi_0&= u^2- \frac{1}{2} -2 \xi_1 -\xi_2 \\
    \xi_3&= \frac{1}{2}-u^2 - \xi_1 -2\xi_2
\end{split}
\ee
Furthermore, we have the following positivity conditions for the probabilities
\be\begin{split}\label{eq:xi_ineq}
0&\leq \tilde q_c(0,0,0) = \tilde q(0,0,0) -\xi_0 = \frac{(u^3+v^3)^2 -r}{2} -\xi_0\\
0&\leq \tilde q_a(1,1,1) = \tilde q(1,1,1) +\xi_3 = \frac{(u^3-v^3)^2+r}{2} +\xi_3
\\  
0&\leq \tilde q_c(0,0,1) = \tilde q(0,0,1) -\xi_1 = \frac{(u^2 v-v^2 u)^2+r}{2} -\xi_1
\end{split}
\ee
We use the first inequality of Eq. \eqref{eq:xi_ineq} to get
\be
\begin{split}
    \xi_0 &\leq \frac{(u^3+v^3)^2 -r}{2}  \\
     u^2- \frac{1}{2} -2 \xi_1 -\xi_2 &\leq \frac{(u^3+v^3)^2 -r}{2}\\
     u^2- \frac{1+(u^3+v^3)^2 -r}{2} -2 \xi_1  &\leq \xi_2.
\end{split}
\ee
The second inequality to bound
\be
\begin{split}
\xi_3 &\geq - \frac{(u^3-v^3)^2+r}{2}\\
\frac{1}{2}-u^2 - \xi_1 -2\xi_2 &\geq - \frac{(u^3-v^3)^2+r}{2} \\
\frac{1+ (u^3-v^3)^2+r}{2} -u^2 -\xi_1 &\geq 2 \xi_2
\end{split}.
\ee
Together the two conditions imply
\be\begin{split} \label{eq:x_1_l}
 2 u^2- (1+(u^3+v^3)^2 -r) -4 \xi_1 \leq   \frac{1+ (u^3-v^3)^2+r}{2} -u^2 -\xi_1 \\
\xi_1 \geq \frac{6 u^2 -2(u^3+v^3)^2 -(u^3-v^3)^2  - 3 +r}{6}
\end{split}
\ee
and the very last inequality of Eq. \eqref{eq:xi_ineq} to simply write
\be \label{eq:x_1_u}
\xi_1 \leq \frac{(u^2 v-v^2 u)^2+r}{2}.
\ee
Combining the equations \eqref{eq:x_1_l} and \eqref{eq:x_1_u} gives 
\be
r \geq R_\theta\equiv \frac{1}{2} \sin ^3(\theta ) (-6 \sin (\theta )+3 \cos (\theta )+\cos (3 \theta ))
\ee
in the notation $u=\cos(\theta)$ and $v=\sin(\theta)$.

\section{Appendix E. Source entanglement and output randomness of the RGB4 distribution. }

For the source states 
\be\label{eq: app state gen}
 \ket{\psi_\xi}= \frac{1}{\sqrt{2}}\big( \ket{01}_{\textbf{X}_\xi \textbf{Y}_\xi}\!\ket{j^{c}_\gamma}_{X'_\xi Y'_\xi E_\xi}\!\!+
\ket{10}_{\textbf{X}_\xi \textbf{Y}_\xi}\ket{j^{a}_\gamma}_{X'_\xi Y'_\xi E_\xi} \big)
\ee
introduce the Schmidt decompositions
\be\label{eq: app junk schmidt}\begin{split}
\ket{j^{c}_\xi}_{X'_\xi Y'_\xi E_\xi} &=\sum_{k} \lambda_{\xi}^{k} \ket{c_{\xi}^{(k)}}_{X'_\xi Y'_\xi} \ket{\kappa_{\xi}^{(k)}}_{E_\xi}\\
\ket{j^{a}_\xi}_{X'_\xi Y'_\xi E_\xi} &=\sum_{k} \mu_{\xi}^{k} \ket{a_{\xi}^{(k)}}_{X'_\xi Y'_\xi} \ket{\sigma_{\xi}^{(k)}}_{E_\xi}
\end{split}
\ee
To shorten the equations we denote the overlaps between Eve's states as $\Gamma_{kj}^\xi = \braket{\sigma_\alpha^{(k)}}{\kappa_\alpha^{(j)}}_{E_\xi}$.
With the explicit parametrization of the junk states, the coherence terms take the form 
\be
\bra{\Psi^a}\Pi^{1_a}_A \Pi^{1_b}_B\Pi^{1_c}_C  \ket{\Psi^c}= \sum_{\substack{k,k',k^\ast \\ j,j',j^\ast}} \mu_\alpha^{k}\mu_\beta^{k'}\mu_\gamma^{k^\ast}
\lambda_\alpha^{j}\lambda_\beta^{j'}\lambda_\gamma^{j^\ast}
\Gamma_{kj}^\alpha \Gamma_{k'j'}^\beta\Gamma_{k^\ast j^\ast }^\gamma
 \bra{\circlearrowleft}^{\otimes 3} \bra{a_\alpha^{(k)} a_\beta^{(k)},a_\gamma^{(k^\ast)} }
\Pi^{1_a}_A \Pi^{1_b}_B\Pi^{1_c}_C\ket{\circlearrowright}^{\otimes 3}\ket{c_\alpha^{(j)},c_\beta^{(j')}c_\gamma^{(j^\ast)}}.
\ee
Where the state $\ket{\circlearrowright}^{\otimes 3}= \ket{10,10,10}_{\textbf{A}_\beta\textbf{A}_\gamma \textbf{B}_\gamma\textbf{B}_\alpha \textbf{C}_\alpha \textbf{C}_\beta} $ denotes the state where all the tokens are sent clockwise, and $\ket{\circlearrowleft}^{\otimes 3}$ where they are sent anti-clockwise.
Note that the two projectors $\{ \Pi_X^{1_0}, \Pi_{X}^{1_1} \}$ define a PVM in in the subspace $\text{span}\{\ket{\cw}_{X_\xi X_\xi'},\ket{\acw}_{X_\xi X_\xi'}\}\otimes \cH_{J_X}$, in which $\ket{\Psi^a}$ and $\ket{\Psi^c}$ belong.

Now consider the coherence $r$ defined in the main text
\be
2 \, \text{Re} \bra{\Psi^a} \Pi^{1_i}_A \Pi^{1_j}_B \Pi^{1_k}_C \ket{\Psi^c} = (-1)^{1+j+k} r,
\ee
and define the following global operator 
\be
V = \sum_{ijk=0,1}  (-1)^{i+j+k} \Pi^{1_i}_A \Pi^{1_j}_B \Pi^{1_k}_C\,
\ee
which is unitary on the subspace where $\ket{\Psi^c}$ and $\ket{\Psi^a}$ belong. We can now compute the coherence 
\be\begin{split}
|\bra{\Psi^a} V \ket{\Psi^c}| &\geq  \text{Re} \bra{\Psi^a} V \ket{\Psi^c} 
= \text{Re} \bra{\Psi_a} \sum_{ijk=0,1}  (-1)^{i+j+k}  \Pi^{1_i}_A \Pi^{1_j}_B \Pi^{1_k}_C \ket{\Psi^c} = \sum_{ijk} \frac{r}{2} = 4\ r.
\end{split}
\ee

On the other hand, with the above parametrization introduced in Eq.~\eqref{eq: app junk schmidt}, can we express it as
\be
\bra{\Psi^a} V \ket{\Psi^c} = \sum_{\substack{k,k',k^\ast \\ j,j',j^\ast}} \mu_\alpha^{k}\mu_\beta^{k'}\mu_\gamma^{k^\ast}
\lambda_\alpha^{j}\lambda_\beta^{j'}\lambda_\gamma^{j^\ast}
\Gamma_{kj}^\alpha \Gamma_{k'j'}^\beta\Gamma_{k^\ast j^\ast }^\gamma
 \bra{\circlearrowleft}^{\otimes 3} \bra{a_\alpha^{(k)} a_\beta^{(k)},a_\gamma^{(k^\ast)} }
V \ket{\circlearrowright}^{\otimes 3}\ket{c_\alpha^{(j)},c_\beta^{(j')}c_\gamma^{(j^\ast)}}.
\ee
A trivial upper bound $| \bra{\circlearrowleft}^{\otimes 3} \bra{a_\alpha^{(k)} a_\beta^{(k)},a_\gamma^{(k^\ast)} }
V \ket{\circlearrowright}^{\otimes 3}\ket{c_\alpha^{(j)},c_\beta^{(j')}c_\gamma^{(j^\ast)}}|\leq 1$ allow us to get 
\be
\begin{split}
4 r \leq |\bra{\Psi^a} V \ket{\Psi^c} | &\leq |\sum_{\substack{k,k',k^\ast \\ j,j',j^\ast}} \mu_\alpha^{k}\mu_\beta^{k'}\mu_\gamma^{k^\ast}
\lambda_\alpha^{j}\lambda_\beta^{j'}\lambda_\gamma^{j^\ast}
\Gamma_{kj}^\alpha \Gamma_{k'j'}^\beta\Gamma_{k^\ast j^\ast }^\gamma| \\
&\leq
|\sum_{kj} \mu_\alpha^k \lambda_\alpha^j \Gamma_{kj}^\alpha| \cdot|\sum_{kj} \mu_\beta^k \lambda_\beta^j \Gamma_{kj}^\beta| \cdot |\sum_{kj} \mu_\gamma^k \lambda_\gamma^j\Gamma_{k j}^\gamma|
\end{split}
\ee

In particular, we use $|\sum_{kj} \mu_\alpha^k \lambda_\alpha^j \Gamma_{kj}^\alpha| \leq 1$ and $|\sum_{kj} \mu_\gamma^k \lambda_\gamma^j\Gamma_{k j}^\gamma| \leq \sum_{kj} \mu_\gamma^k \lambda_\gamma^j| \Gamma_{k j}^\gamma|$ to get a bount useful for the next section
\be\label{bi}
\left(\sum_{kj} \mu_\beta^k \lambda_\beta^j |\Gamma_{kj}^\beta| \right) \left(  \sum_{kj} \mu_\gamma^k \lambda_\gamma^j|\Gamma_{k j}^\gamma|  \right) \geq 4 r
\ee
\subsection{Randomness}

To quantify the randomness produced by the measurement we focus on the conditional entropy of a single output, say $a$, with respect to an eavesdropper. For simplicity we further coarse-grain the values of $a$ to define a bit $\bar a =0$ (for $a=0,2$) and $\bar a = 1$ (if $a=1_0,1_1$), encoded in the register $\bar A$. We are  interested in the classical-quantum state $\varrho_{\bar A E}$ shared between the register $\bar A$ and the eavesdropper. The outcomes $\bar{a}$ are determined by the states of the systems $\textbf{A}_\beta$ and $\textbf{A}_\gamma$, which are only correlated with $E_\beta$ and $E_\gamma$. For the state of interest, we obtain  
\be\begin{split}
\varrho_{\bar A E} &= \frac{1}{4} \ketbra{0}_{\bar A} \left(\rho^{(c)}_{ E_\beta}\otimes \rho^{(a)}_{ E_\gamma} + \rho^{(a)}_{ E_\beta}\otimes \rho^{(c)}_{ E_\gamma}\right) \\ 
&+ \frac{1}{4} \ketbra{1}_{\bar A} \otimes  \left(\rho^{(c)}_{ E_\beta}\otimes \rho^{(c)}_{ E_\gamma} + \rho^{(a)}_{ E_\beta}\otimes \rho^{(a)}_{ E_\gamma}\right), 
\end{split}
\ee
where the eavesdropper's states are
\be\label{eq app: rho Eve}
\begin{split}
 \rho^{(c)}_{ E_\xi} &= \sum_k (\lambda_\xi^{(k)})^2 \ketbra{\kappa_\xi^{(k)}}\\
  \rho^{(a)}_{ E_\xi} &= \sum_k (\mu_\xi^{(k)})^2 \ketbra{\sigma_\xi^{(k)}}.
\end{split}
\ee
The randomness of $\bar a$ can be quantified by the min-entropy $H_\text{min}(\bar A|E)= -\log_2{P_\text{guess}(\bar A|E)}$, where $P_\text{guess}(\bar A|E)$ is the probability that Eve guesses the value $\bar a$ correctly. It is related  by $P_\text{guess}(\bar A|E)= \frac{1}{2}\big(1+ D(\rho_{E|\bar a=0},\rho_{E|\bar a=1})\big)$ to the trace distance $D$ between the conditional states of Eve
\be\begin{split}
\rho_{E|\bar a=0} &=\frac{1}{2}\left(\rho^{(c)}_{ E_\beta}\otimes \rho^{(a)}_{ E_\gamma} + \rho^{(a)}_{ E_\beta}\otimes \rho^{(c)}_{ E_\gamma}\right)\\
\rho_{E|\bar a=1} &=\frac{1}{2}\left(\rho^{(c)}_{ E_\beta}\otimes \rho^{(c)}_{ E_\gamma} + \rho^{(a)}_{ E_\beta}\otimes \rho^{(a)}_{ E_\gamma}\right).
\end{split}
\ee
In turn, the trace distance $D\leq \sqrt{1-F^2}$ is bounded by the fidelity $F(\rho,\sigma)=\tr |\sqrt{\rho} \, \sqrt{\sigma}|$ between the states, which gives us
\be
H_\text{min}(\bar A|E)\geq -\log_2\left(\frac{1}{2}(1+\sqrt{1-F^2(\rho_{E|\bar a=0},\rho_{E|\bar a=1})}) \right)
\ee

Using the strong convexity of $F$ we obtain 
\be\label{eq: app sum F}
F(\rho_{E|\bar a=0},\rho_{E|\bar a=1}) \geq \frac{1}{2} \left(F(\rho^{(c)}_{ E_\beta},\rho^{(a)}_{ E_\beta})+ F(\rho^{(c)}_{ E_\gamma},\rho^{(a)}_{ E_\gamma})\right).
\ee
Then we can use the strong convexity again together with the Eqs.~\eqref{eq app: rho Eve} to obtain 
\be
F(\rho^{(c)}_{ E_\xi},\rho^{(a)}_{ E_\xi}) \geq \sum_{k,j} \lambda_\xi^{k} \mu_\xi^{j}|\braket{\kappa_\xi^{(k)}}{\sigma_\xi^{(j)}}| = \sum_{kj} \lambda_\xi^{k} \mu_\xi^{j} |\Gamma_{kj}^\xi|.
\ee
Therefore 
\be
F(\rho_{E|\bar a=0},\rho_{E|\bar a=1}) \geq \frac{1}{2}\left(\sum_{kj} \lambda_\beta^{k} \mu_\beta^{j} |\Gamma_{kj}^\beta| +\sum_{kj} \lambda_\gamma^{k} \mu_\gamma^{j} |\Gamma_{kj}^\gamma| \right).
\ee
And it remains to lower bound the right hand side, given the constaint $\left(\sum_{kj} \mu_\beta^k \lambda_\beta^j |\Gamma_{kj}^\beta| \right) \left(  \sum_{kj} \mu_\gamma^k \lambda_\gamma^j|\Gamma_{k j}^\gamma|  \right) \geq 4 r$  derived in the previous section (Eq.~\ref{bi}). Denoting $X =\sum_{kj} \lambda_\beta^{k} \mu_\beta^{j} |\Gamma_{kj}^\beta|$ and $Y =\sum_{kj} \lambda_\gamma^{k} \mu_\gamma^{j} |\Gamma_{kj}^\gamma|$ the problem is simply
\be
\begin{split}
    \min_{X,Y} \, &\frac{1}{2}(X+Y) \\
    \text{such that}\,   &X \cdot Y \geq 4\, r 
\end{split}
\ee
Without big surprise the minimum is attained at $ X= Y =\sqrt{4 r}$. This allows us to conclude 
\be
F(\rho_{E|\bar a=0},\rho_{E|\bar a=1}) \geq \sqrt{ 4 r},
\ee
and finally get the desired bound on the min entropy
\be
H_\text{min}(\bar A|E) \geq -\log_2 \left(\frac{1}{2}(1-\sqrt{1-4 r}) \right).
\ee
At the optimal value $\theta$ this given about $4\%$ of a bit.


\subsection{Entanglement}

Assume that the entanglement of formation of $\rho^{(\alpha)}$ is at most $\cE_F$. Then one can decompose this state as
\be\label{eq: decomp app}
\rho^{(\alpha)} = \sum p_k \ketbra{\bar{\psi}_\alpha^{(k)}},
\ee
where each $\bar{\psi}_\alpha^{(k)}$ has entanglement entropy of entanglement $S_k$, and $\sum_k p_k S_k \leq \cE_F$. By Eq.~\eqref{eq: rigid states} we known $\rho^{(\alpha)}$ and each $\bar{\psi}_\alpha^{(k)}$ are supported in the subspace projected by $(\ketbra{01} + \ketbra{10} )_{\textbf{B}_\alpha \textbf{C}_\alpha}$. Hence,  each state in the decomposition is of the form
\be
\ket{\bar{\psi}_\alpha^{(k)}}= c_k \ket{01}_{\textbf{B}_\alpha\textbf{C}_\alpha}  \ket{\phi_k}_{B'_\alpha C'_\alpha} + s_k e^{\ii\varphi_k} \ket{10}_{\textbf{B}_\alpha\textbf{C}_\alpha}  \ket{\zeta_k}_{B'_\alpha C'_\alpha},
\ee
with some real positive $c_k$ and $s_k$ positive  satisfying $c_k^2+s_k^2=1$ by normalization. The entropy of entanglement of each state is lower bounded by 
\be
S_k = S\left(tr_{\textbf{B}_\alpha B'_\alpha} \ketbra{\bar{\psi}_\alpha^{(k)}}\right) \geq S\left(tr_{\textbf{B}_\alpha B'_\alpha C'_\alpha} \ketbra{\bar{\psi}_\alpha^{(k)}}\right) = 
S\left(\left(\begin{array}{cc}
     c_k^2&  \\
     & s_k^2
\end{array}\right)\right) \equiv h_\text{bin}(c_k^2)
\ee

Now recall that $\rho^{(\alpha)} = \tr_{E_\alpha} \ketbra{\psi_\alpha}$, hence the decomposition in Eq.~\eqref{eq: decomp app} can be purified, implying for the original state
\be\begin{split}
\ket{\psi_\alpha} &= \sum_k \sqrt{p_k} \ket{\bar{\psi}_\alpha^{(k)}}_{\textbf{B}_\alpha\textbf{C}_\alpha B'_\alpha C'_\alpha} \ket{k}_{E_\alpha}\\
& =\frac{1}{\sqrt{2}}\ket{01}_{\textbf{B}_\alpha\textbf{C}_\alpha} \underbrace{\left(\sqrt{2}\sum_k \sqrt{p_k}  c_k \ket{\phi_k}_{B'_\alpha C'_\alpha} \ket{k}_{E_\alpha} \right)}_{=\ket{j_\alpha^c}_{J_\alpha}} + \frac{1}{\sqrt{2}} \ket{10}_{\textbf{B}_\alpha\textbf{C}_\alpha} \underbrace{\left(\sqrt{2}\sum_k \sqrt{p_k} s_k e^{\ii \varphi_k} \ket{\zeta_k}_{B'_\alpha C'_\alpha} \ket{k}_{E_\alpha} \right)}_{=\ket{j_\alpha^a}_{J_\alpha}}
\end{split}
\ee
with orthogonal states $\ket{k}_{E_\alpha}$.

The bound $|\bra{\Psi^a} V \ket{\Psi^c}|\geq 4r$  implies
\be
\begin{split}
    4 r &\leq |\bra{\Psi^a} V \ket{\Psi^c}|\\
    &= |\bra{10,10,10}_{\textbf{B}_\alpha \textbf{C}_\alpha \textbf{C}_\beta \textbf{A}_\beta  \textbf{A}_\gamma  \textbf{B}_\gamma}\bra{j_\alpha^a,j_\beta^a,j_\gamma^a}_{J_\alpha J_\beta J_\gamma}V\ket{01,01,01}_{\textbf{B}_\alpha \textbf{C}_\alpha \textbf{C}_\beta \textbf{A}_\beta  \textbf{A}_\gamma  \textbf{B}_\gamma}\ket{j_\alpha^c,j_\beta^c,j_\gamma^c}_{J_\alpha J_\beta J_\gamma}|\\
    &\leq \sum_k p_k 2 |c_k s_k| |\bra{10,10,10}\bra{\zeta_l}_{B'_\alpha C'_\alpha}\bra{j_\beta^a,j_\gamma^a} V \ket{10,10,10}\ket{\phi_l}_{B'_\alpha C'_\alpha}\bra{j_\beta^c,j_\gamma^c} |\\
    & \leq 2\sum_k p_k  |c_k s_k|.
\end{split}
\ee
The coherence value thus guarantees $\sum_{k} p_k |c_k s_k| \geq 2 r$.

Now let use denote $c_k^2=q_k$ and minimize the average entropy of entanglement under the constraint imposed by the coherence. Formally, it amounts to solve the minimization problem
\be\begin{split}
\min_{q_k, p_k} &\sum_k p_k h_\text{bin}(q_k )\\
    \text{such that }\, & \sum_{k} p_k \sqrt{q_k(1-q_k)}\geq 2 r,
\end{split}
\ee
where each $q_k$ is between zero and one, and $p_k$ define a probability distribution. To solve it we define new variables $R_k = \sqrt{q_k(1-q_k)} \in[0,1/2]$, with $q_k=\frac{1}{2} \left(1-\sqrt{1-4 R_k^2}\right)$, and solve 
\be\begin{split}
\min_{R_k, p_k} &\sum_k p_k h_\text{bin}\left(\frac{1-\sqrt{1-4 R_k^2}}{2}\right)\\
    \text{such that }\, & \sum_{k}p_k  R_k\geq 2 r.
\end{split}
\ee
One can verify that $\frac{\dd ^2 h_\text{bin}\left(\frac{1-\sqrt{1-4 R_k^2}}{2}\right)}{\dd R^2} \geq 0$, so the goal function is convex. It is thus minimized by setting all $R_k =2 r$. Therefore for any decomposition (Eq.~\ref{eq: decomp app}) the average entropy of entanglement satisfies $\sum_{k}p_k h_\text{bin}(q_k)\geq h_\text{bin}\left(\frac{1-\sqrt{1-4(2r)^2}}{2}\right)$. By definition the same bound holds for the entanglement of formation of the mixed state $\rho^{(\alpha)}$
\be
\cE_F \geq h_\text{bin}\left(\frac{1-\sqrt{1-16 r^2}}{2}\right).
\ee
For the optimal value one finds $\cE_F \geq 2.5 \%$.

\section{Appendix F. Self-testing Parity Token Counting distributions on the triangle}

In the PTC strategy on the triangle network, each source $S_\alpha$, $S_\beta$, and $S_\gamma$ have a single token. The source $\xi$ sends the token to the left  with probably $p_\xi$ and to the right  with probability $1- p_\xi$. Each party outputs  the \textit{parity} $a,b, c \in \{0,1\}$ of the number of received tokens. By construction, one has
\be\label{app eq: PTC1}
a \oplus b\oplus c = 1 
\ee
with $\oplus$ denoting addition modulo 2. A PTC distribution can also be obtained from any TC counting strategy if each party only outputs the parity of the total number of received tokens. 

Classical rigidity of PTC distributions on the triangle was shown in~\cite{boreiri2022} for all distributions steaming from strategies with $p_{\alpha},p_{\beta},p_{\gamma} \neq \frac{1}{2}$. Whenever a distribution corresponds to  $p_k=\frac{1}{2}$ for at least one of the sources, it turns out that it can be simulated with a whole family of nonequivalent PTC strategies. However, in this case, rigidity holds, but only up to this freedom to choose the nonequivalent PTC strategy; see \cite{boreiri2022} for details.

Here we extend this result to quantum models. In the PTC task, the measurements have binary outputs and correspond to the PVMs $\{\Pi_X^0,\Pi_X^1\}$. 
Now, similar to the TC case, let us define a unitary operator for each party X as following
\be
U_X = \Pi^1_X - \Pi^0_X.
\ee
Eq.~\eqref{app eq: PTC1} guarantees that 
\be
U_AU_BU_C \ket{\Psi} = \ket{\Psi},
\ee
from which, by result 1, we conclude that all the unitaries are product 
\be\nonumber\begin{split}
U_{A} &= V_{A_\beta}\otimes W_{A_\gamma}\\
U_{B} &= V_{B_\gamma}\otimes W_{B_\alpha}\\
U_{C} &= V_{C_\alpha}\otimes W_{A_\beta}.    
\end{split}
\ee

Without loss of generality we can chose the eigenvalues of $V_{A_\beta}$ and $W_{A_\gamma}$ such that $V_{A_\beta} = \Pi^{1}_{A_\beta}-\Pi^{0}_{A_\beta}$ and $W_{A_\gamma} = \Pi^{1}_{A_\gamma}-\Pi^{0}_{A_\gamma}$, which guarantees
\be\begin{split}
\Pi^1_A &= \Pi^{1}_{A_\beta}\otimes \Pi^{1}_{A_\gamma} +  \Pi^{0}_{A_\beta}\otimes \Pi^{0}_{A_\gamma} = \sum_{a_\beta \oplus a_\gamma = 0} \Pi^{a_\beta}_{A_\beta}\otimes \Pi^{a_\gamma}_{A_\gamma} \\
\Pi^0_A &= \Pi^{1}_{A_\beta}\otimes \Pi^{0}_{A_\gamma} +  \Pi^{0}_{A_\beta}\otimes \Pi^{1}_{A_\gamma} = \sum_{a_\beta \oplus a_\gamma = 1} \Pi^{a_\beta}_{A_\beta}\otimes \Pi^{a_\gamma}_{A_\gamma}.    
\end{split}
\ee
This allows us to split the Hilbert decompose  associated to each system as $\cH_{A_\beta} = \mathds{C}^2_{\textbf{A}_\beta}\otimes \cH_{J_{A\beta}}$ in such a way that

\be\begin{split}
\Pi_A^a &= \left( \sum_{j \oplus \ell = a} \ketbra{j,\ell}\right)_{\textbf{A}_\beta \textbf{A}_\gamma} \otimes \mathds{1}_{J_{A\beta} J_{A\gamma}}\\
\ket{\psi_\alpha}_{B_\alpha C_\alpha}&=\sum_{i,j \in \{0,1\}}  \Psi_{ij}^{(\alpha)} \ket{i,j}_{\textbf{B}_\alpha \textbf{C}_\alpha}\ket{j^{(ij)}_\alpha}_{J_{B\alpha}J_{C\alpha}}.
\end{split}
\ee

Note that this form corresponds to a classical strategy with the response function $a=j\oplus \ell$ that is explicitly PTC. The same decomposition can be guaranteed for each party and each source. It remains to show that the amplitudes $\Psi_{ij}^{(\alpha)}$ (that can be taken real by absorbing the phase inside the junk states) are enforced to be unique. As in the TC case, this follows from the classical rigidity of PTC distributions -- which guarantee that for the generic case with $p_{\alpha},p_{\beta},p_{\gamma} \neq \frac{1}{2}$, the probabilities $P^{(\alpha)}(i,j) = |\Psi_{ij}^{(\alpha)}|^2$ are essentially unique upon relabeling $0$ and $1$, see ~\cite{boreiri2022}. Therefore, we have

\be 
\ket{\psi_\alpha}_{B_\alpha C_\alpha}= \sqrt{p_{\alpha}} \ket{01}_{\textbf{B}_\alpha \textbf{C}_\alpha} \ket{j^c_\alpha}_{J_{B\alpha}J_{C\alpha}} + ~~
\sqrt{1-p_{\alpha}}
\ket{10}_{\textbf{B}_\alpha \textbf{C}_\alpha} \ket{j^a_\alpha}_{J_{B\alpha}J_{C\alpha}}
\ee

which illustrates the self-testing (or quantum rigidity) result, similar to result 2, for PTC distributions on the triangle.

\end{widetext}
\end{document}